\documentclass[showpacs,pre,a4paper,11pt,groupedaddress]{revtex4}
\usepackage{graphicx}
\usepackage{epsfig}
\usepackage{amsmath,amssymb}

\begin{document}

\title{Microfield distributions in strongly coupled two-component plasmas}
\author{H.~B.~Nersisyan}
\altaffiliation{Permanent address: Institute of Radiophysics and Electronics, 378410
Ashtarak, Armenia}
\email{nersisyan@theorie2.physik.uni-erlangen.de}
\author{C.~Toepffer}
\author{G.~Zwicknagel}
\affiliation{Institut f\"ur Theoretische Physik II, Erlangen-N\"urnberg Universit\"at,
Staudtstrasse 7, D-91058 Erlangen, Germany}
\date{\today}

\begin{abstract}
The electric microfield distribution at charged particles is studied
for two-component electron-ion plasmas using molecular dynamics simulation and
theoretical models. The particles are treated within classical statistical mechanics
using an electron-ion Coulomb potential regularized at distances less than the de
Broglie length to take into account the quantum-diffraction effects. The
potential-of-mean-force (PMF) approximation is deduced from a canonical ensemble
formulation. The resulting probability density of the electric microfield satisfies
exactly the second-moment sum rule without the use of adjustable parameters. The
correlation functions between the charged radiator and the plasma ions and electrons
are calculated using molecular dynamics simulations and the hypernetted-chain
approximation for a two-component plasma. It is shown that the agreement between the
theoretical models for the microfield distributions and the simulations is quite good
in general.
\end{abstract}

\maketitle

\section{Introduction}

Because of the Stark effect, the fluctuating electric microfields created by
the charged particles in a plasma influence its optical and thermodynamic
properties. They affect the profiles of spectral lines (broadening and
shift) and effectively lower the photoionization thresholds of atoms and ions
immersed in a plasma \cite{1,2,3}. A comparison of experimental and
theoretical widths and shapes of the Stark-broadened spectral lines is widely
used for plasma diagnostics \cite{4,5}.

Under certain assumptions \cite{1,2}, the observed spectral line shapes can be
closely related to the electric microfield distribution at the radiating atom
or ion (radiator) \cite{6,7}. Within the quasistatic approximation the problem
is then reduced to a determination of the probability distribution of the
low-frequency component of the perturbing electric fields. This is mainly
associated with the distribution of the heavier perturbing particles, i.e. the
ions, whereas the electrons can be assumed to adjust instantaneously to the
configuration of the ions.

Since the pioneering work of Holtsmark \cite{6}, who completely neglected
correlations between the particles (ideal plasma), many efforts have been
concentrated on an improved statistical description of the microfield
distribution. The first theory which goes beyond the Holtsmark limit and
which is based on a cluster expansion similar to that of Ursell and Mayer
\cite{8} was developed by Baranger and Mozer \cite{9,10}. In this approach
the microfield distribution is represented as an expansion in terms of
correlation functions which has been truncated on the level of the pair
correlation. The latter is treated in the Debye-H\"uckel form which
corresponds to the first order of the expansion in the coupling parameter.
The theory by Baranger and Mozer was improved by Hooper \cite{11,12} and
later by Tighe and Hooper \cite{13,14}. Based on Broyles' collective-coordinate
technique \cite{15} they reformulated the expansion of the microfield distribution
in terms of other functions by introducing a free parameter which was adjusted
in such a way to arrive at a level where the resulting microfield distribution
did not depend on the free parameter any more. A further improvement of this
model was made in Ref.~\cite{16} considering a Debye-chain cluster expansion.
Afterwards the Baranger-Mozer second order theory was extended by including
higher order corrections, like the triple correlation contribution \cite{17,18}.
However, it was argued that such a method is only valid for low-density,
high-temperature plasmas, i.e. at small coupling parameters, where the correction
to the Holtsmark distribution, corresponding to the first term in the series,
is small. In the limit of very strong coupling and without screening Meyer's
harmonic oscillator model is applicable \cite{19}, in which every ion is assumed
to oscillate independently of the others around its equilibrium position
at the ion-sphere center. The first theory capable to provide reliable
numerical results for strongly coupled plasmas, known as adjustable-parameter
exponential approximation (APEX), was proposed by Iglesias, Lebowitz \textit{et al.}
\cite{20,21,22,23,24}. This phenomenological but highly successful approximation
is based on a special parameterization of the electric microfield produced
on a radiator. It involves a non-interacting quasiparticle representation of the
electron-screened ions, designed to yield the correct second moment of the
microfield distribution. APEX was first developed for three- and two-dimensional
Coulomb systems \cite{21,22} and later adapted to screened Coulomb systems and
ion mixtures \cite{23,24}. (See also Ref.~\cite{25} for the corrected version
of APEX for a neutral radiator). Another approach providing reliable numerical
results for the strongly coupled plasmas was proposed by Iglesias \cite{26}.
Following the idea of Morita \cite{27} on the similarity of the representation
of the microfield distribution to that of the excess chemical potential, Iglesias
reduced the problem to a determination of the radial distribution function (RDF)
for a fictitious system with an imaginary part in the interaction energy. Employing
this idea Lado and Dufty \cite{28,29,30} developed an integral equation technique
for calculating the RDF and good agreement was found with computer simulations.
It is now possible to calculate the microfield distribution from Monte Carlo (MC)
or molecular-dynamics (MD) simulations of plasmas \cite{31,32,33,34,35}. These methods
allows to study the effects of microfield nonuniformity \cite{36,37} and the dynamical
properties of the electric microfield \cite{35,38} as well as to simulate the
high-frequency microfield distribution in electron plasmas \cite{35,39}. With these
powerful tools one can check the accuracy of theoretical models and establish
asymptotic or analytic fitting formulas suitable for applications (see, e.g.
\cite{40,41} and references therein).

Until now most work was done on either electronic or ionic one-component plasmas
(OCP) neglecting the influence of the attractive interactions between electrons
and ions. Here we treat ions and electrons on an equal footing by concentrating
on two-component plasmas (TCP). Previously this has been done in Ref.~\cite{42}
for partially degenerate electrons. In particular, the low-frequency component
of the microfield was calculated within the linear response treatment taking strong
correlations into account via local field corrections. Also the problem of attractive
interaction has been considered for single but highly charged impurity ion immersed
in an electronic OCP (see, e.g. Ref.~\cite{43} for a recent review of these cases).

In the present paper we study strongly coupled systems, i.e. a highly charged
radiator in a TCP of classical (nondegenerated) and strongly correlated particles
beyond a perturbative treatment. As in Ref.~\cite{42} the presented theoretical
scheme is based on the potential-of-mean-force (PMF) approximation which exactly
satisfies the sum-rule requirement arising from the second moment of the microfield
distribution without introducing adjustable parameters. Another important ingredient
is the electron-ion attractive interaction which drastically changes the physical
properties of the system as compared to classical OCPs (see, e.g., \cite{43}).
This may cause significant changes in the microfield distribution on either neutral
or charged radiators. But the thermodynamic stability of a TCP requires some
quantum features for the electron-ion interaction at short distances. Here we
focus on an application of classical statistical mechanics and MD simulations
which is enabled by using a regularized ion-electron potential where the divergence
at the origin is removed \cite{27,44}, see also \cite{43,45} for a review.

The paper is organized as follows. In Sec.~II, we define the systems and parameters
of interest as well as the theoretical model to calculate the microfield distribution
in a TCP. The exact second moment for the charged radiator is
calculated in Sec.~III. The theoretical schemes applied previously to either
electronic or ionic OCPs are generalized to TCPs in Sec.~IV. In particular,
we consider the Holtsmark distribution, express the microfield distribution through
the pair distribution functions, and deduce the PMF approximation from the classical
canonical ensemble. Furthermore we construct a theoretical approach based on the
exponential approximation where the effective electric fields are calculated on the
basis of the PMF approximation and the pair correlation functions.
In Sec.~V we consider the hypernetted-chain (HNC) integral equations technique to
calculate these functions in a two component plasma. In order to test the theoretical
models we carried out classical MD simulations to calculate both the pair correlation
functions and the microfield distribution. Technical aspects and the numerical
results are presented in Sec.~V. These results are summarized in Sec.~VI. Some
details of the calculations are described in the Appendix.

\section{Microfield distribution in a TCP: Theoretical background}

\subsection{Basic parameters for the TCP}

We consider a neutral and isotropic two component electron-ion plasma consisting
of $N_{i}$ ions and $N_{e}$ electrons at a temperature $T$ in a volume $\Omega$.
The particles are assumed to be classical and pointlike. The average densities,
charges and masses of the ions and electrons are $n_{i}=N_{i}/\Omega $,
$n_{e}=N_{e}/\Omega$, and $Ze$, $-e$ and $m_{i}$, $m$, respectively. We assume that
the density of radiator ions are small, $n_{R}\ll n_{i;e}$ and thus consider only
one radiator ion with charge $Z_{R}e$ in our calculations (throughout this paper
the index $R$ refers to the radiators). Because of the charge neutrality we have
$N_{i}Z-N_{e}+Z_{R}=0$. In the thermodynamic limit ($N_{i;e}\rightarrow
\infty$ and $\Omega\rightarrow\infty$) this is equivalent to $n_{e}=n_{i}Z$.

We now introduce the Coulomb coupling parameters $\Gamma_{\alpha\beta}$ which play
an important role for characterizing the properties of a TCP. Introducing the
Wigner-Seitz radii, i.e. the mean electron-electron, electron-ion and ion-ion
distances through the relations, $a_{e}^{-3}=4\pi n_{e}/3$, $a^{-3}=4\pi n/3$ and
$a_{i}^{-3}=4\pi n_{i}/3$ (where $n=n_{e}+n_{i}$ is the plasma total density) these
parameters are defined as
\begin{equation}
\Gamma_{ee}=\frac{e_{S}^{2}}{a_{e}k_{B}T}, \quad
\Gamma_{ei}=\frac{Ze_{S}^{2}}{ak_{B}T}, \quad
\Gamma_{ii}=\frac{Z^{2}e_{S}^{2}}{a_{i}k_{B}T},
\end{equation}
respectively, where $e_{S}^{2}=e^{2}/4\pi\varepsilon_{0}$. Note that
\begin{equation}
\Gamma_{ee}=\frac{\Gamma_{ei}}{\left[Z^{2}\left(Z+1\right)\right] ^{1/3}},
\quad \Gamma_{ii}=\frac{Z\Gamma_{ei}}{\left(Z+1\right)^{1/3}}.
\end{equation}
In a hydrogen plasma with $Z=1$ we obtain $\Gamma_{ee}=\Gamma_{ii}=2^{-1/3}%
\Gamma_{ei}$ while in a plasma with highly charged ions ($Z\gg 1$) $%
\Gamma_{ii}=Z^{2/3}\Gamma_{ei}$ and $\Gamma_{ee}=\Gamma_{ei}/Z$. For $%
Z\geqslant 2$ the coupling parameters satisfy the inequality $%
\Gamma_{ee}<\Gamma_{ei}<\Gamma_{ii}$.

Here we consider the pair interaction potential $e_{S}^{2}q_{\alpha}q_{\beta}%
u_{\alpha\beta}\left(r\right)$ with $\alpha;\beta =e,i,R$, $q_{e}=-1$, $q_{i}=Z$,
$q_{R}=Z_{R}$, and
\begin{equation}
u_{\alpha \beta }\left( r\right) =\frac{1}{r}\left( 1-e^{-r/\delta _{\alpha
\beta }}\right)
\end{equation}%
which is regularized at small distances due to quantum-diffraction effects.
In this paper we assume that the Coulomb potential is cutoff
at the thermal de Broglie wavelengths, $\delta_{\alpha\beta}=\left(\hbar^{2}/%
\mu_{\alpha\beta}k_{B}T\right)^{1/2}$, where $\mu_{\alpha\beta}$ is the reduced mass
of the particles $\alpha$ and $\beta$. For large distances $r>\delta_{\alpha\beta}$
the potential becomes Coulomb, while for $r<\delta_{\alpha\beta}$ the Coulomb
singularity is removed and $u_{\alpha\beta}(0)=1/\delta_{\alpha\beta}$. By this
the short range effects based on the uncertainty principle are included
\cite{27,43,44,45}.

For a classical description of a plasma the electron degeneracy parameter
$\Theta_{e}$, i.e. the ratio of the thermal energy and the Fermi energy must
fulfill $\Theta_{e}=k_{B}T/E_{F}>1$. Or, alternatively, the electron thermal
wavelength should be smaller than the electron-electron mean distance, $\delta_{ee}<2\left(4/9\pi\right)^{1/3} a_{e}\simeq 1.04a_{e}$. Since an ion
is much heavier than an electron this condition is usually fulfilled
for ions. We note that $\delta_{ii}\ll\delta_{ei}$ and $\delta_{ee}\simeq%
2^{1/2}\delta_{ei}$ since $\mu_{ei}\simeq m$. Therefore one can expect that the
regularization given by Eq.~(3) is less important for ions than for
electrons. Furthermore, scattering of any two particles is classical for
impact parameters that are large compared to the de Broglie wavelengths.
Typical impact parameters are given by the Landau lengths, $\lambda_%
{L\alpha\beta}=e_{S}^{2}\left|q_{\alpha}q_{\beta}\right|/k_{B}T$. Its ratio
to the de Broglie wavelengths is given by
\begin{equation}
\sigma_{\alpha\beta}=\frac{\lambda_{L\alpha\beta}}{\delta_{\alpha\beta}}=
\Gamma_{ei}\frac{\left|q_{\alpha}q_{\beta}\right|}{Z}\frac{a}{\delta_{\alpha%
\beta}}=\frac{e_{S}^{2}\left|q_{\alpha}q_{\beta}\right|u_{\alpha\beta}\left(0%
\right)}{k_{B}T}.
\end{equation}
This is also the maximum value of the interparticle interaction energy in the
units of $k_{B}T$, where $\sigma_{ee}<\sigma_{ei}\ll \sigma_{ii}$ and $\sigma_{ei}%
\simeq 2^{1/2}Z\sigma_{ee}$. Classical description of the scattering events
in the TCP is valid if $\sigma_{ee}>1$. This can be alternatively
written in the explicit form $k_{B}T<1\,$Ry.
Combining this condition with the one considered above we finally obtain the
temperature domain where the classical treatment is adequate, $E_{F}<k_{B}T<1\,$Ry.
This condition occurs at lower densities of electrons. Since the parameter $\delta_{ee}$
increases with electron-ion Coulomb coupling the classical condition $\sigma_{ee}>1$
implies that the state with stronger $\Gamma_{ei}$ behaves more classical as discussed
in Ref.~\cite{43}.

\subsection{Microfield distribution formulation within thermodynamic
canonical ensemble}

The electric microfield distribution (MFD) $Q\left(\boldsymbol%
{\varepsilon}\right)$ is defined as the probability density of finding a field
${\bf E}=\boldsymbol{\varepsilon}$ at a charge $Z_{R}e$, located
at ${\bf r}_{0}$, in a TCP with $N_{i}$ ions and $N_{e}$ electrons. This system
is described by classical statistical mechanics in a canonical ensemble of
$\left(N_{i}+N_{e}+1\right)$ particles, and temperature $T$. The normalized
probability density of the microfield $\boldsymbol{\varepsilon}$
in the thermodynamic limit is then given by
\begin{equation}
Q\left(\boldsymbol{\varepsilon}\right) =\frac{1}{W}\int_{\Omega }
e^{-\beta_{T}U\left({\cal T}_{e},{\cal T}_{i},\mathbf{r}_{0}\right)} \delta\left(%
\boldsymbol{\varepsilon}-\mathbf{E}\left({\cal T}_{e},{\cal T}_{i},\mathbf{r}%
_{0}\right) \right) d\mathbf{r}_{0}d{\cal T}_{e}d{\cal T}_{i},
\end{equation}%
where $\beta_{T}=1/k_{B}T$, and ${\cal T}_{e}=\left\{{\bf r}_{1},{\bf r}_{2}...{\bf r}_{N_{e}}
\right\}$, ${\cal T}_{i}=\left\{{\bf R}_{1},{\bf R}_{2}...{\bf R}_{N_{i}}\right\}$
are the coordinates of electrons and ions, respectively. Here
\begin{equation}
W=\int_{\Omega }e^{-\beta_{T}U\left({\cal T}_{e},{\cal T}_{i},\mathbf{r}%
_{0}\right)} d\mathbf{r}_{0}d{\cal T}_{e}d{\cal T}_{i}
\end{equation}%
is the canonical partition function and $U\left({\cal T}_{e},{\cal T}_{i},%
\mathbf{r}_{0}\right)$ is the potential energy of the configuration
\begin{equation}
U\left({\cal T}_{e},{\cal T}_{i},\mathbf{r}_{0}\right) =U_{ee}\left({\cal T}_{e}\right)+U_{ii}\left({\cal T}_{i}\right) +U_{ei}\left({\cal T}_{e},{\cal T}_{i}\right) +U_{eR}\left({\cal T}_{e},\mathbf{r}_{0}\right)
+U_{iR}\left({\cal T}_{i},\mathbf{r}_{0}\right)
\end{equation}%
with electron-electron, ion-ion, electron-ion, electron-radiator and ion-radiator
interaction terms, respectively. Assuming spherical symmetric interactions
between the particles the interaction terms in Eq.~(7) can be represented as
\begin{equation}
U_{\alpha\beta}\left({\cal T}_{\alpha},{\cal T}_{\beta}\right)=\vartheta
_{\alpha\beta }q_{\alpha}q_{\beta}e_{S}^{2}\sum_{a,b}u_{\alpha\beta
}\left(\left\vert\mathbf{r}_{a}^{\left(\alpha\right)}-{\bf r}_{b}^{\left%
(\beta\right)}\right\vert\right),
\end{equation}
\begin{equation}
U_{\alpha R}\left({\cal T}_{\alpha},{\bf r}_{0}\right)=q_{\alpha }Z_{R}e_{S}^{2}
\sum_{a}u_{\alpha R}\left(\left\vert{\bf r}_{0}-{\bf r}_{a}^{\left(\alpha\right)}
\right\vert\right)
\end{equation}%
in terms of the pair interaction potentials $u_{\alpha \beta }\left( r\right)$
and $u_{\alpha R}\left(r\right)$, where $\alpha;\beta=e;i$, $\vartheta
_{ee}=\vartheta _{ii}=1/2$, $\vartheta _{ei}=1$, $\mathbf{r}_{a}^{\left(
e\right) }=\mathbf{r}_{a}$, $\mathbf{r}_{a}^{\left( i\right) }=\mathbf{R}%
_{a} $. In Eq.~(8) the sum is restricted to $a\neq b$ for like particles, $\alpha=
\beta$. The total electrical field $\mathbf{E}\left({\cal T}_{e},{\cal T}_{i},%
\mathbf{r}_{0}\right) $ acting on the radiator is given by the superposition
of electronic and ionic single-particle fields
\begin{equation}
\mathbf{E}\left({\cal T}_{e},{\cal T}_{i},\mathbf{r}_{0}\right) =-\frac{1}{%
Z_{R}e}\mathbf{\nabla }_{0}U=\mathbf{E}_{e}\left({\cal T}_{e},\mathbf{r}%
_{0}\right) +\mathbf{E}_{i}\left({\cal T}_{i},\mathbf{r}_{0}\right)
\end{equation}%
with
\begin{equation}
\mathbf{E}_{\alpha}\left({\cal T}_{\alpha},\mathbf{r}_{0}\right)
=\sum_{a=1}^{N_{\alpha }}\mathbf{E}_{\alpha }\left( \mathbf{r}_{0}-\mathbf{r}%
_{a}^{\left( \alpha \right) }\right) .
\end{equation}%
As ${\bf E}_{e}\left({\bf r}\right)=\frac{{\bf r}}{r}E_{e}\left%
(r\right)$, ${\bf E}_{i}\left({\bf r}\right)=\frac{{\bf r}}{r}E_{i}\left(r\right)$,
we obtain for the electronic and ionic single-particle fields $E_{e}\left(r\right)
=e_{F}u_{eR}^{\prime}\left(r\right)$, $E_{i}\left(r\right) =-Ze_{F}u_{iR}^{\prime}\left(r\right)$, where the prime indicates
derivative with respect to $r$, and $e_{F}=e/4\pi\varepsilon_{0}$.

The spherical symmetric interaction between plasma particles allows to introduce
the normalized microfield distribution $P(\varepsilon)=4\pi\varepsilon^{2}%
Q(\varepsilon)$. It is useful to consider the Fourier transform of
$Q\left(\boldsymbol{\varepsilon}\right)$ defined by
\begin{equation}
T\left({\bf K}\right)=\int Q\left(\boldsymbol{\varepsilon}\right)%
e^{i{\bf K}\cdot \boldsymbol{\varepsilon}}d\boldsymbol{\varepsilon}
=\left<e^{i{\bf K}\cdot {\bf E}}\right>.
\end{equation}
Here $\left<...\right>$ denotes a statistical average. Again we note that due to
the isotropy of the system the Fourier transform of the MFD must behave as
\begin{eqnarray}
T\left(K\right)=\int_{0}^{\infty }P\left(\varepsilon\right) j_{0}\left(
K\varepsilon\right) d\varepsilon, \quad P\left(\varepsilon\right)=\frac{%
2\varepsilon^{2}}{\pi}\int_{0}^{\infty }T\left(K\right) j_{0}\left(K
\varepsilon\right) K^{2}dK,
\end{eqnarray}
where $j_{0}(x)=\sin x/x$ is the spherical Bessel function of order zero. The
coefficients of the expansion of the function $T(K)$ at $K\rightarrow 0$
yield the even moments of the microfield distribution,
\begin{equation}
T\left(K\right)=1-\frac{K^{2}}{6}\left<E^{2}\right>+\frac{K^{4}}
{120}\left<E^{4}\right> -...
\end{equation}
The similar expansion for the function ${\cal L}\left(K\right)$ defined by
$T(K)=e^{-{\cal L}(K)}$ yields
\begin{equation}
{\cal L}\left(K\right)=\frac{K^{2}}{6}\left<E^{2}\right>+\frac{K^{4}}{72}
\left[\left<E^{2}\right>^{2}-\frac{3}{5}\left<E^{4}\right>%
\right]+...
\end{equation}
Therefore the Fourier transform of the MFD can be interpreted as a generating
function for microfield even moments. Moreover, Eqs.~(14) and (15) suggest a
simple criterion for the existence of even moments. In particular, the second moment
of the MFD exists if the function ${\cal L}\left(K\right)$ and its first
and second derivatives are regular at the origin. Eqs.~(5)-(15) then describe the
total MFD at the position ${\bf r}_{0}$ of the radiator generated by both the
statistically distributed ions and electrons of the TCP.
Since we are interested to calculate the MFD, Eq.~(5), in an infinite system
the statistical average of any quantity becomes translationally invariant with
respect to ${\bf r}_{0}$, and the location of the test charge may be taken as the
origin without loss of generality.

\section{Second moment}

A knowledge of moment sum rules is often useful in developing approximation
schemes for fluids and plasmas. The moments of the MFD fix the shape
of the distribution and involves some useful information about the system.
For example, the exact second moment has been previously incorporated into
the calculation of the MFDs in the APEX scheme. Here, we derive exact expressions
for the second moment of the MFD on charged radiators. Note that in general the
existence of the second moment requires that the MFD decays at large electric fields
faster than $\varepsilon^{-3}$.

Let us consider the exact expression for the second moment of the
microfield distribution in the TCP and for a charged radiator. The second
moment may be written in the form
\begin{equation}
\left<E^{2}\right>=\frac{1}{\left(Z_{R}e\right)^{2}}\left<\left(\boldsymbol{\nabla}%
_{0}U\right)^{2}\right>,
\end{equation}
where $\boldsymbol{\nabla}_{0}$ is the gradient with respect to ${\bf r}_{0}$ and
the average is over the canonical ensemble defined in Eq.~(5). Noting that $%
e^{-\beta_{T}U}\left(\boldsymbol{\nabla}_{0}U\right)=-k_{B}T\left(\boldsymbol{%
\nabla }_{0}e^{-\beta_{T}U}\right)$, substituting this relation into
Eq.~(16), integrating by parts, and setting the surface terms equal to zero
yields
\begin{equation}
\left<E^{2}\right>=\frac{k_{B}T}{\left(Z_{R}e\right)^{2}}\left<\nabla_{0}%
^{2}U\right>=-\frac{k_{B}T}{Z_{R}e}\left<\left(\boldsymbol{\nabla}_{0}\cdot
{\bf E}\right)\right>.
\end{equation}%
We now use Eqs.~(10) and (11), the relation $\boldsymbol{\nabla}\cdot{\bf
E}_{\alpha}({\bf r})=\left(q_{\alpha}e_{F}/r^{2}\right)\widetilde{u}_{\alpha}(r)$,
where $\widetilde{u}_{\alpha}\left(r\right)=-\left[r^{2}u_{\alpha R}^{\prime}%
\left(r\right)\right]^{\prime}$, and translational symmetry. This yields
\begin{equation}
\left<E^{2}\right>=\frac{k_{B}Tn_{e}}{Z_{R}\varepsilon_{0}}\left[
\int_{0}^{\infty }\widetilde{u}_{e}\left( r\right) g_{eR}\left( r\right)
dr-\int_{0}^{\infty }\widetilde{u}_{i}\left( r\right) g_{iR}\left( r\right)
dr\right].
\end{equation}
The functions $g_{\alpha R}(r)$ are the pair correlation functions between radiator
and the plasma particles, where $n_{\alpha}g_{\alpha R}(r)$ is the density of plasma
particles $\alpha$ at a distance $r$ from the radiator. These functions can be
represented as
\begin{equation}
g_{eR}\left(r_{1}\right)=\frac{\Omega^{2}}{W}\int_{\Omega}e^{-\beta_{T}U\left({\cal
T}_{e},{\cal T}_{i}\right)} d{\cal T}^{(1)}_{e} d{\cal T}_{i},
\end{equation}
\begin{equation}
g_{iR}\left( R_{1}\right) =\frac{\Omega^{2}}{W}\int_{\Omega }e^{-\beta_{T}U\left({\cal
T}_{e},{\cal T}_{i}\right)} d{\cal T}_{e}d{\cal T}_{i}^{(1)}.
\end{equation}%
Here $d{\cal T}^{(s)}_{\alpha}=\prod_{a=s+1}^{N_{\alpha}} d{\bf r}^{(\alpha)}_{a}$
is the reduced
volume element in a phase space which does not involve the particles $1,2,...s$
of plasma species $\alpha$. The interaction potential energy, $U\left({\cal T}_{e}%
,{\cal T}_{i}\right)$, does not depend on ${\bf r}_{0}$. The pair correlation
functions given by Eqs.~(19) and (20) describe the coupling between radiator ion
and plasma particles. For a vanishing radiator-plasma coupling, e.g. for a neutral
radiator the pair correlation functions behave like $g_{\alpha R}\rightarrow 1$. If
the radiator is a particle of plasma species $\beta$ these correlations functions
coincide with the radial distribution functions (RDF) of bulk plasma, $g_{\alpha R}%
\equiv g_{\alpha\beta}$.

The second moment for the regularized Coulomb interaction (see Eq.~(3)) is with
$\widetilde{u}_{\alpha }\left( r\right)=\left( r/\delta _{\alpha R}^{2}\right)
e^{-r/\delta _{\alpha R}}$
\begin{equation}
\left<E^{2}\right>=\frac{k_{B}Tn_{e}}{Z_{R}\varepsilon_{0}}\left[\frac{1%
}{\delta _{eR}^{2}}\int_{0}^{\infty }e^{-r/\delta _{eR}}g_{eR}\left(
r\right) rdr-\frac{1}{\delta _{iR}^{2}}\int_{0}^{\infty }e^{-r/\delta
_{iR}}g_{iR}\left( r\right) rdr\right] .
\end{equation}
Using a bare Coulomb interaction $\widetilde{u}_{\alpha }\left( r\right)=\delta\left(
r\right)$ in Eq.~(18) one recovers the result obtained in Ref.~\cite{42}
\begin{equation}
\left<E^{2}\right>=\frac{k_{B}Tn_{e}}{Z_{R}\varepsilon_{0}}\left[
g_{eR}\left(0\right)-g_{iR}\left(0\right)\right],
\end{equation}%
which can also be obtained from Eq.~(21) by taking the limits $\delta
_{eR}\rightarrow 0$, $\delta_{iR}\rightarrow 0$. For $Z_{R}>0$, we may
assume that $g_{iR}\left(0\right)=0$ if quantum-diffraction effects
are negligible for the ions, while $g_{eR}\left(r\right)$ diverges at small
distances for
a bare Coulomb potential. This indicates that the second moment of the microfield
distribution does not exist for a classical Coulomb TCP. But in the OCP limit
$g_{eR}\left(
0\right)=1$ one recovers the result $\left<E^{2}\right>_{\rm OCP}=k_{B}TZn_{i}/Z_{R}
\varepsilon_{0}$ for the classical (ionic) OCP \cite{21}.

\section{Approximate calculations of the MFD}

In this section we generalize the existing theoretical approaches developed
originally for a OCP to a two component electron-ion plasma. For practical
applications we will consider the exponential approximation
considered in Ref.~\cite{46}, and, as a simple but useful example the Holtsmark
limit for the MFD in a TCP.

\subsection{Ideal plasmas: Holtsmark distribution}

We first consider the microfield distribution in an ideal TCP with
$\Gamma_{ee},\Gamma_{ei},\Gamma_{ii}\rightarrow 0$, i.e. in the high temperature
regime $T\rightarrow \infty$. In this case Eq.~(12) yields
\begin{equation}
T\left(K\right)=\prod_{\alpha}\left\{ 1-\frac{4\pi n_{\alpha}}{N_{\alpha}}
\int_{\Omega}\left[1-j_{0}\left(KE_{\alpha}\left(r\right)\right)\right]
r^{2}dr\right\}^{N_{\alpha}}.
\end{equation}
In the thermodynamic limit ($N_{\alpha}$,~$\Omega\rightarrow\infty$,
$N_{\alpha}/\Omega=n_{\alpha}={\rm const}$), and recalling that
$T(K)=e^{-{\cal L}(K)}$ we obtain from Eq.~(23)
\begin{equation}
{\cal L}\left(K\right)=\sum_{\alpha}4\pi n_{\alpha}\int_{0}^{\infty }
\left[ 1-j_{0}\left(KE_{\alpha}\left(r\right)\right)\right] r^{2}dr.
\end{equation}
We study this expression for two types of interaction potentials.

(i)~For a bare Coulomb interaction Eq.~(24) yields ${\cal L}\left(
K\right)=\left(KE_{H}\right)^{3/2}$, where $E_{H}$ is the Holtsmark field
for a TCP, $E_{H}^{3/2}=E_{He}^{3/2}+E_{Hi}^{3/2}$. Here $E_{He}$ and $E_{Hi}$ are
the electronic and ionic Holtsmark fields, respectively, $%
E_{He}=Ce_{F}/a_{e}^{2}$, $E_{Hi}=CZe_{F}/a_{i}^{2}$ with $C=\left( 8\pi /25\right)
^{1/3}$. Since $E_{He}=Z^{-1/3}E_{Hi}$ the electronic and ionic components of
a hydrogen TCP contribute equally to the Holtsmark field. For a completely
ionized TCP with highly charged ions the ions dominate $E_{H}$. The definition
of the Holtsmark field $E_{H}$ for a TCP is equivalent to the obvious
relation $n=n_{e}+n_{i}$ and can be represented as
\begin{equation}
E_{H}=\left(\frac{8\pi}{25}\right)^{1/3}\frac{{\cal Z}e_{F}}{a^{2}}
=\left(\frac{8\pi}{25}\right)^{1/3}\frac{e_{F}}{a^{2}}
\left[ \frac{Z\left( 1+Z^{1/2}\right) }{Z+1}\right] ^{2/3}
\end{equation}
with an effective charge ${\cal Z}$.
For a hydrogen TCP with $Z=1$ also ${\cal Z}=1$. In other cases the effective
charge increases with $Z$ and behaves as ${\cal Z}=Z^{1/3}$ for large $Z$.
Thus the ideal two-component plasma can be regarded as an ionic OCP with effective
ionic charge ${\cal Z}$.

Since the function ${\cal L}(K)$ has a singularity at $K=0$ it cannot be
expanded there and the second moment does not exist. The microfield distribution
is given by $P_{H}\left(E\right) =H\left(\eta\right)/E_{H}$ in terms of Holtsmark's
function $H(\eta)$
\begin{equation}
H\left(\eta\right)=\frac{2\eta}{\pi}\int_{0}^{\infty }e^{-x^{3/2}}\sin
\left(\eta x\right)xdx
\end{equation}
with $\eta =E/E_{H}$. Note that the Holtsmark distribution for the TCP has the same
functional form as either the ionic or the electronic OCP. The only difference is the
definition of the Holtsmark field. Since the electronic or ionic Holtsmark fields may
significantly differ from $E_{H}$ the shape of the MFD for a OCP and a TCP may strongly
differ from each other even for ideal plasmas.

(ii) For the regularized Coulomb interaction given by Eq.~(3),
${\cal L}(K)$ (from Eq.~(24)) and all its derivatives are regular at $K=0$. Hence,
all moments of the microfield
distribution exist. This indicates that for large electric fields the
microfield distribution must decay exponentially. The second moment
can be obtained from Eq.~(24) if we recall that for $K\rightarrow 0$,
${\cal L}(K)\simeq\left(K^{2}/6\right)\left<E^{2}\right>$, thus
\begin{equation}
\left<E^{2}\right>=2\pi n_{e}e_{F}^{2}\left(\frac{1}{\delta_{eR}}+\frac{Z}{%
\delta _{iR}}\right).
\end{equation}
For large electric fields the main contribution to the microfield distribution
comes from small $K$ and we obtain the asymptotic behavior
\begin{equation}
P(E)\simeq 3\sqrt{\frac{6}{\pi}}\frac{E^{2}}{\left<E^{2}\right>^{3/2}}\exp\left(-%
\frac{3E^{2}}{2\left<E^{2}\right>}\right),
\end{equation}
where $\left<E^{2}\right>$ is given by Eq.~(27). For large $K$ ($K\rightarrow%
\infty$) the function ${\cal L}(K)$ (Eq.~(24)) behaves as for the bare Coulomb interaction
${\cal L}\left(K\right)\simeq\left(KE_{H}\right)^{3/2}$. Hence, the microfield
distributions for the ideal plasmas with bare and regularized Coulomb potentials behave
similar at small electric fields.

\subsection{Expression of the MFD through pair functions}

It was first noted by Morita \cite{27} that the virial expansion of the Fourier
transform of the MFD $T({\bf K})$ is formally similar to that of the
excess chemical potential. This was previously used to express
$T({\bf K})$ in terms of an effective RDFs (see, e.g., \cite{26})
involving the radiator and one of the plasma particles. To generalize this method
to the TCP we follow the procedure [6-30] and consider the logarithmic
derivative of Eq.~(12)
\begin{equation}
-\frac{\partial{\cal L}\left({{\bf K}}\right)}{\partial K}=
i\frac{\left<\left(\hat{{{\bf K}}}\cdot {\bf E}\right)
e^{i{{\bf K}}\cdot {\bf E}}\right>}{\left<e^{i{{\bf K}}
\cdot {\bf E}}\right>}=i\hat{{\bf K}}\cdot\sum_{\alpha} n_{\alpha}
\int d{\bf r}{\bf E}_{\alpha}\left({\bf r}\right)\left[{\cal G}_{\alpha R}%
\left({\bf r},{{\bf K}}\right)-1\right].
\end{equation}
Here ${\bf E}_{e}\left({\bf r}\right)$ and ${\bf E}_{i}\left({\bf r}\right)$
are the single-particle electronic and ionic electrical fields introduced above, and
$\hat{{{\bf K}}}$ is a unit vector in the direction of ${{\bf K}}$.
${\cal G}_{eR}\left({\bf r},{{\bf K}}\right)$ and ${\cal G}_{iR}
\left({\bf r},{{\bf K}}\right)$ represent the pair correlation
functions between the radiator and the plasma particles in a fictitious system
whose interaction potential is given by the complex quantity ${\cal U}\left
({\cal T}_{e},{\cal T}_{i},{\bf K}\right)=U({\cal T}_{e},{\cal T}_{i})-
i\left(k_{B}T\right)\left({\bf K}\cdot{\bf E}\right)$, i.e.,
\begin{eqnarray}
{\cal G}_{eR}\left({\bf r}_{1},{\bf K}\right)&=&\frac{\Omega^{2}}{%
{\cal W}({\bf K})}\int_{\Omega}e^{-\beta_{T}{\cal U}\left({\cal T}_{e},{\cal T}_{i},{\bf K}\right)}d{\cal T}_{e}^{(1)}d{\cal T}_{i}, \\
{\cal G}_{iR}\left({\bf R}_{1},{\bf K}\right)&=&\frac{\Omega^{2}}{%
{\cal W}({\bf K})}\int_{\Omega}e^{-\beta_{T}{\cal U}\left({\cal T}_{e},{\cal T}_{i},{\bf K}\right)}d{\cal T}_{e}d{\cal T}^{(1)}_{i}
\end{eqnarray}%
with the generalized, reduced partition function ${\cal W}({\bf K})
\equiv {\cal W}(K)=W\left\langle e^{i{{\bf K}}\cdot
\mathbf{E}}\right\rangle$. It can be easily checked that this function is real.
In general these correlation functions are complex and satisfy the
symmetry relations ${\cal G}_{\alpha R}\left(-{\bf r},{{\bf K}}\right)
={\cal G}_{\alpha R}^{\ast}\left({\bf r},{{\bf K}}\right)$ and ${\cal
G}_{\alpha R}\left(-{\bf r},-{{\bf K}}\right)={\cal G}_{\alpha R}\left
({\bf r},{{\bf K}}\right)$, where the asterix denotes the complex conjugate.
The correlation functions in the fictitious system are not spherical
symmetric. At ${\bf K}=0$ they coincide with $g_{\alpha R}(r)$ given by Eqs.~(19)
and (20). The complex correlation functions ${\cal G}_{\alpha R}$ can be expressed
through two functions ${\cal G}^{(0)}_{\alpha R}\left(r,K\right)$ and
$\boldsymbol{\cal E}^{(0)}_{\alpha}\left({\bf r},K\right)=\hat{\bf r}{\cal
E}^{(0)}_{\alpha}\left(r,K\right)$,
\begin{equation}
{\cal G}_{\alpha R}\left({\bf r},{{\bf K}}\right)={\cal G}^{(0)}_{\alpha
R}\left(r,K\right) \exp\left[i{\bf K}\cdot \boldsymbol{\cal
E}^{(0)}_{\alpha}\left({\bf r},K\right)\right],
\end{equation}
where ${\cal G}^{(0)}_{\alpha R}\left(r,K\right)$ and ${\cal E}^{(0)}_{\alpha}
\left(r,K\right)$ are spherical symmetric real functions. With this choice the functions
${\cal G}_{\alpha R}\left({\bf r},{\bf K}\right)$ automatically satisfy the symmetry
relations. In the limit ${\bf K}\rightarrow 0$, we have also ${\cal G}^{(0)}_{\alpha
R}\left(r,0\right)=g_{\alpha R}(r)$. Inserting Eq.~(32) into Eq.~(29), integrating over
$K$ and taking into account that ${\cal L}(0)=0$ we obtain
\begin{equation}
{\cal L}\left(K\right)=4\pi\sum_{\alpha}n_{\alpha}\int_{0}^{\infty}E_{\alpha}
\left(r\right) r^{2}dr\int_{0}^{K}{\cal G}^{(0)}_{\alpha R}\left(r,\lambda
\right)j_{1}\left(\lambda{\cal E}^{(0)}_{\alpha}\left(r,\lambda\right)\right)d\lambda,
\end{equation}
where $j_{1}\left(x\right)=-j_{0}^{\prime }\left(x\right)$. Eq.~(33) is an exact
result which allows to express the MFD through complex pair correlation functions
(or, alternatively through two real functions). In addition Eq.~(33) yields the exact
second moment given by Eq.~(18).

The problem is now the evaluation of these correlation functions. Eq.~(33) requires
that the complex correlation functions has to be known in
the interval from $0$ to $K$. One possibility is to apply the integral equation
technique with the complex interaction energy introduced above. Such an approach has
been previously employed for a OCP \cite{28,29,30} and shows good agreement with
computer simulations. Here we adopt the exponential approximation (see, e.g., \cite{21,22,23,24,25,26,46}) and generalize it to the TCP. This method is based on thermodynamic perturbation theory
\cite{47}. The system with the potential energy ${\cal U}(K=0)=U$ is chosen
as reference system and its structure is assumed to be known to a good approximation.
The perturbation potential is then given by $U_{1}=-i\left(k_{B}T\right)\left(
{\bf K}\cdot {\bf E}\right)$ and we expand the correlation functions,
Eqs.~(30) and (31), with respect to $U_{1}$. Within first order we obtain
${\cal G}_{\alpha R}\left({\bf r},{\bf K}\right) \simeq g_{\alpha R}
\left(r\right)\left[1+i{\bf K}\cdot\boldsymbol{{\cal E}}_{\alpha}\left(
{\bf r}\right) \right] $. Here $g_{\alpha R}\left(r\right)$ are the actual RDF in the
real system, Eqs.~(19) and (20), and $\boldsymbol{{\cal E}}_{\alpha}
\left({\bf r}\right)=\boldsymbol{{\cal E}}^{(0)}_{\alpha}\left({\bf r},0\right)$.
The electric fields $\boldsymbol{\cal E}_{\alpha }\left({\bf r}\right)$ may be
interpreted as effective electric fields in the fictitious system which are
independent of ${\bf K}$. Taking into account that $\left<{\bf E}\right>=0$
we obtain
\begin{equation}
\boldsymbol{\cal E}_{\alpha}\left(\mathbf{r}\right)=\mathbf{E}_{\alpha}\left(%
\mathbf{r}\right) +\frac{1}{g_{\alpha R}\left(r\right)}\sum_{\beta}n_{\beta}%
\int d\mathbf{r}_{1}\mathbf{E}_{\beta}\left(\mathbf{r}_{1}\right) \left[%
g_{\alpha\beta}\left(\left\vert\mathbf{r}-\mathbf{r}_{1}\right\vert\right)-1%
\right].
\end{equation}
Comparing Eq.~(34) with Eq.~(3.8) of Ref.~\cite{21} for the case of a OCP we remark
that our present derivation yields an additional factor $1/g_{\alpha R}(r)$ in front
of the second term.
Since $g_{\alpha\beta}(r)$ depend only on $\left\vert{\bf r}_{1}-{\bf
r}_{2}\right\vert$ the effective electric fields in Eq.~(34) can be
represented as ${\boldsymbol{\cal E}}_{\alpha}\left({\bf r}\right)=\hat{\bf r}
{\cal E}_{\alpha}(r)$. The ${\cal E}_{\alpha}(r)$ can be expressed by the pair
correlation functions and the single-particle potentials $u_{\alpha R}\left(r\right)$
(see Appendix~A for details). Alternatively the Fourier transformed single-particle
electric fields can be written as $\mathbf{E}_{\alpha}\left( \mathbf{k}\right)
=\hat{\bf k}E_{\alpha}(k)$ which allows to express the effective fields through
the static structure factors $S_{\alpha\beta}\left(k\right)$.

We make now the ansatz,
\begin{equation}
{\cal G}_{\alpha R}\left({\bf r},{\bf K}\right)=g_{\alpha R}\left(
r\right)\exp\left[i{\bf K}\cdot\boldsymbol{{\cal E}}_{\alpha }\left(
{\bf r}\right) \right]
\end{equation}
and then integrate Eq.~(33) with respect to $\lambda$, to find
\begin{equation}
{\cal L}\left(K\right)=\sum_{\alpha}4\pi n_{\alpha}\int_{0}^{\infty}E_{\alpha}%
\left(r\right) \frac{1-j_{0}\left(K{\cal E}_{\alpha}\left(r\right)\right)}%
{{\cal E}_{\alpha}\left(r\right)}g_{\alpha R}\left(r\right) r^{2}dr.
\end{equation}
The second moment within the exponential approximation can be found from Eq.~(36)
at $K\rightarrow 0$ and results in
\begin{equation}
\left<E^{2}\right>=\sum_{\alpha} 4\pi n_{\alpha}\int_{0}^{\infty}E_{\alpha}%
\left(r\right) {\cal E}_{\alpha}\left(r\right) g_{\alpha R}\left(r\right)%
r^{2}dr.
\end{equation}%
This must fulfill the exact second moment of the MFD given by Eq.~(18) or (33),
which is not affected by either the assumption~(35) or its first order Taylor
expansion with respect to ${\bf K}$.

The APEX approach was originally developed for the classical ionic OCP with
bare Coulomb interaction. In order to fulfill the exact second moment
$\left<E^{2}\right>_{\rm OCP}=k_{B}TZn_{i}/Z_{R}\varepsilon_{0}$, Eq.~(37) must
take the form
\begin{equation}
\int_{0}^{\infty} {\cal E}\left(r\right) g_{R}\left(r\right) dr=\frac{k_{B}T}{%
Z_{R}e}.
\end{equation}
In Ref.~\cite{21} the effective field ${\cal E}(r)$ is assumed as a Debye-H\"uckel
like screened interaction with unknown screening length. This free parameter
is then adjusted in such a way to satisfy Eq.~(38). The resulting predictions of APEX
for the probability densities show excellent agreement with numerical simulation data
for the OCP. However, difficulties appear when one attempts to extend the APEX scheme
to a TCP, e.g. by assuming
a Debye-H\"uckel like interaction separately for the electrons and the ions and
introducing two adjustable screening lengths. Then the sum rule Eq.~(38)
with the exact second moment $\left<E^{2}\right>$ becomes ambiguous as it allows
for many different choices of the adjustable screening
lengths. This can be cured for ionic mixtures by demanding that the second moment rule
is satisfied species by species (see, e.g., \cite{23,24}). But this cannot
be employed for a TCP with attractive electron-ion interactions. Here the Debye-H\"uckel
ansatz for the electronic effective field is physically incompatible with Eq.~(37) as
discussed in Ref.~\cite{42}.

We instead apply the potential of mean force (PMF) approximation
\cite{42,48} which expresses the effective electric fields through the logarithmic
derivative of pair correlation functions
\begin{equation}
{\cal E}_{\alpha}\left(r\right)=\frac{k_{B}T}{Z_{R}e}\frac{\partial}{\partial r}%
\left[\ln g_{\alpha R}\left(r\right)\right].
\end{equation}
Introducing Eqs.~(39) in Eq.~(37) automatically satisfies the sum-rule~(18)
without any adjustable parameter. Relations~(39) can
be deduced from Eqs.~(19), (20) and (34). To show this we consider
the pair correlation functions given by Eqs.~(19) and (20). It is clear that in the
thermodynamic limit these expressions are translationally invariant with respect to
${\bf r}^{(\alpha)}_{a}\rightarrow {\bf r}^{(\alpha)}_{a}+{\bf r}^{(\alpha)}_{1}$
($a=2,3,...,N_{\alpha}$). Making these transformations and calculating the
logarithmic derivatives of the pair correlation functions yield Eq.~(39), where
the effective fields are given by Eq.~(34). In addition, Eq.~(39) can
be interpreted as a integro-differential equation for determining the pair correlation
functions, $g_{\alpha R}$. Thus, if the $g_{\alpha R}(r)$ are known the MFD with the
exact second moment can be calculated using Eqs.~(13), (36) and
(39). This approach based on the exponential and the PMF approximations is
abbreviated as PMFEX in the following.

We summarize this section by the following remarks. The possibilities of the PMF
approximation have already been noted by Alastuey \textit{et al.} \cite{22}.
They found a superiority of the APEX to the PMF approximation since the former
reproduces the simulation data for classical ionic OCP more accurately than the
latter. We have confirmed this by own investigations on the OCP. For the TCP the
outlined PMFEX approximation agrees quite well with the MD simulation results, as
we will show in the next section.

\section{Results}

In Sec.~IV we introduced and outlined the PMFEX approximation which links the MFD
to the RDFs. To obtain explicit results for the MFD the corresponding RDFs have to
be determined first. This will be done by solving numerically the Hyper-Netted-Chain
(HNC) integral equations for the TCPs under consideration. The HNC method and the
PMFEX approximation are tested both by comparison of the resulting  RDFs and MFD with
those obtained by classical MD simulations. We have done that for a wide range of coupling
parameters $\Gamma_{ei}$ and for two specific rather distinct cases H$^{+}$ ($n_{e}=%
n_{i}$) and Al$^{13+}$ ($n_{e}=13n_{i}$) TCPs with symmetric and asymmetric density
distributions between plasma species, respectively. For simplicity we assume bare
Coulomb electron-electron and ion-ion interactions with $\delta_{ee}\simeq 0$ and
$\delta_{ii}\simeq 0$ while the parameter $\delta_{ei}/a=\bar\delta$ scaled in the
Wigner-Seitz radius $a=\left[4\pi\left(n_{e}+n_{i}\right)/3\right]^{-1/3}$ varies
from $0.1$ to $0.4$.

\subsection{Numerical treatments}

To determine the RDFs $g_{\alpha R}(r)$ the HNC equations (see, e.g., Refs.~\cite{48,49})
\begin{equation}
1+h_{\alpha\beta}\left(r\right)=\exp\left[h_{\alpha\beta}\left(r\right)
-c_{\alpha\beta}\left(r\right)-\beta_{T}q_{\alpha}q_{\beta}e^{2}_{S}u_%
{\alpha\beta}(r)\right],
\end{equation}
and the Ornstein-Zernike equations
\begin{equation}
h_{\alpha \beta }\left( r\right) =c_{\alpha \beta }\left( r\right)
+\sum_{\sigma }n_{\sigma }\int d\mathbf{r}^{\prime }c_{\alpha \sigma }\left(
\left\vert \mathbf{r}-\mathbf{r}^{\prime }\right\vert \right) h_{\sigma
\beta }\left( r^{\prime }\right)
\end{equation}
for the total correlation functions $h_{\alpha\beta}\left(r\right)=g_{\alpha\beta} \left(r\right)-1$ and the direct correlation functions $c_{\alpha\beta}(r)$ are
considered. This has to be done for a three-component system of electrons, ions and
the radiator in general. Here we assume that the radiator is one of the plasma ions
$Z_{R}=Z$, i.e.~$g_{\alpha R}(r)\rightarrow g_{\alpha i}(r)$, which reduces Eqs.~(40)
and (41) to the HNC-scheme for a TCP with mutual interactions $u_{\alpha\beta}(r)$
(see Eq.~(3)). The resulting coupled equations~(40) and (41) are solved numerically
by an iterative scheme which closely follows the implementation discussed in detail
in Ref.~\cite{50}. Within our numerical treatment a parameter regime with
$\sigma_{ei}=\sigma=\Gamma_{ei}/\bar\delta<\sigma_{c}(Z,{\bar\delta})$ is accessible,
where the critical value $\sigma_{c}$ for ${\bar\delta}=0.1$,~$0.2$~and~$0.4$ takes the
values $\sigma_{c}\simeq 8.32$,~$8.5$,~$13.4$ and $\sigma_{c}\simeq 7.33$,~$6.66$,~$7.0$
for H$^{+}$ and Al$^{13+}$ TCPs, respectively.
Beyond this value the numerical procedure does either not converge or ends up in
unphysical solutions. A similar behavior has been reported in \cite{43} for the case
of an ion embedded in electrons. With the RDFs provided by the HNC-scheme the MFD,
i.e.~$P(E)$, is then calculated via Eqs.~(13), (36) and (39) by standard numerical
differentiation and integration methods \cite{numrep}.

In the MD simulations the classical equations of motion are numerically integrated for
$N_i$ ions and $N_e$ electrons interacting via $u_{\alpha\beta}(r)$ and contained
in a cubic cell with periodic boundary conditions. To account for the long range of the
Coulomb interaction the forces are calculated by an Ewald sum \cite{nij57,han73}.
The numerical propagation is accomplished by a standard Velocity-Verlet algorithm
\cite{ver67,all87} extended by a hierarchical treatment of close colliding particles
which are propagated as subsystems (see \cite{zwi99} for details) and using an adaptive
time step. Such MD simulations have already been extensively tested and successfully
applied for investigations of the dynamic response of a TCP with regularized potentials,
see \cite{psch03,zwi03}.

The actual simulations run with $N=N_{i}+N_{e}=2002$ particles and proceed in two
phases. An initial equilibration starts from a random sampling of positions and
velocities and relaxes towards the equilibrium distribution of desired temperature
by dynamic propagation with velocity rescaling. The subsequent simulations are
performed in the microcanonical ensemble, where their accuracy and stability can be
monitored using the total energy. The MFD and the RDFs are sampled during the simulations
from the known forces on the particles and their positions as a time average over the
total running time $\tau$ which was typically $\tau\approx 700 \omega_{\rm pl,e}^{-1}$,
where $\omega_{\rm pl,e} = (n_e e^2/m \epsilon_0)^{1/2}$ is the electronic plasma
frequency.

By the MD simulations basically all correlations and many-body effects of classical
many-body systems can be taken into account. Limitations arise mainly from the finite
particle number and the system size, e.g. in connection with the screening of the
interactions on a typical screening length $\lambda_D$. Since $\lambda_D$ should be
smaller than the size $L$ of the simulation box and $L/\lambda_D \propto \Gamma_{ei}^%
{1/2}$, the MD-technique works here more favorable at large coupling ($\Gamma_{ei}>1$)
while the limit of weak coupling ($\Gamma_{ei}\ll 1$) requires a strong increase of
the simulation box, i.e. of the particle number.

\subsection{Correlation functions}

In Figs.~1-3 we compare the RDFs calculated either from the HNC scheme or MD
simulations for a H$^{+}$-plasma. Only $g_{ei}(r)$ and $g_{ii}(r)$ are plotted
as $g_{ee}(r)=g_{ii}(r)$ for hydrogen both in the HNC and MD treatment (within
numerical fluctuations). In Fig.~1 we explore the dependence in the regularization
parameter $\delta$ at a fixed electron-electron coupling $\Gamma_{ee}=0.1$, while
in Fig.~2 $\Gamma_{ee}$ is varied. Both approaches agree perfectly in the range
of parameters covered in these figures. Due to the regularization of the ion-electron
interaction the RDF $g_{ei}(r)$ is finite in the limit $r\rightarrow 0$. A non-linear
Debye-H\"uckel approximation for $g_{ei}(r)$ has been proposed in Ref.~\cite{43}.
Adopting this estimate for TCPs we obtain $g_{ei}(0)\simeq \exp\left(\Gamma_{ei}/%
{\cal R}{\bar\delta}\right)$ with ${\cal R}=1+{\bar\delta}\left(3\Gamma_{ei}%
\right)^{1/2}$, where the dependence on the ion charge $Z$ is included in the
coupling parameter $\Gamma_{ei}$. The RDFs show indeed the expected growth of
correlations with increased coupling and decreased regularization parameter.
For very strong electron-electron coupling deviations between the HNC scheme and
the MD simulations begin to appear as shown in Fig.~3 for $\Gamma_{ee}=4.0$ and
$\sigma_{{\rm H}^{+}}\simeq 12.5$. There we are at the edge of the HNC convergence
region and classically bound states show up in the MD simulations.

The symmetry between the correlation functions $g_{ee}(r)$ and $g_{ii}(r)$ breaks
down for Al$^{13+}$-plasma. In Fig.~4 we compare the HNC and MD radial distribution
functions for fixed $\Gamma_{ee}=0.01$ varying the regularization parameter $\delta$.
Since $\Gamma_{ii}\gg \Gamma_{ee}$ strong correlation effects are expected for $g_{ii}$.
The increasing "correlation hole" is clearly visible in Fig.~5 where the HNC and MD
radial distribution functions $g_{ii}(r)$ and $g_{ei}(r)$ are plotted for a fixed
$\bar\delta=0.4$ and varying coupling strengths $\Gamma_{ee}$. Again, for all these
parameters the HNC scheme agrees perfectly with the MD simulations. As shown in
Fig.~6 deviations occur in the electron-electron RDF $g_{ee}(r)$ at small $r$ for
strong coupling $\Gamma_{ee}=0.2$, $\sigma_{{\rm Al}^{13+}}\simeq 6.6$. These are
due to the enhancement of the electronic density around an ion, which also increases
the probability of close electronic distances and results in the maxima in $g_{ee}$
at distances $r\lesssim a$. This effect is obviously overestimated in the HNC approach
and it is more pronounced for highly charged ions like Al$^{13+}$ and less important
for H$^{+}$. The regularization of the electron-ion interaction has no visible
influence on the correlation functions $g_{ee}(r)$ and $g_{ii}(r)$ (see Figs.~1~and~4).

\subsection{Microfield distribution}

We now turn to the MFDs at the charged reference point which is chosen to be one
of the plasma ions, $Z_{R}=Z$. For our analysis it is instructive to consider first
the second moment which can be used to check and compare the different treatments
PMFEX, HNC, MD, and can provide some information about the shape of the MFD, although
this is not sufficient to construct it. For a bare ion-ion Coulomb interaction the
second term in the rhs of Eq.~(21) vanishes and the second moment $\left<E^{2}\right>$
receives contribution only from the first term involving $g_{ei}$. In the limit of
an ideal TCP ($\Gamma_{\alpha\beta}\rightarrow 0$) $g_{ei}(r)$ can be replaced by
unity. This yields $\left<E^{2}\right>_{0}=k_{B}Tn_{e}/Z\varepsilon_{0}=\left%
(3/\Gamma_{ii}\right)E_{0i}^{2}=\left(3/Z^{5/3}\Gamma_{ee}\right)E_{0i}^{2}$ with
$E_{0i}=Ze_{F}/a_{i}^{2}$ which is similar to the second moment obtained for the
ionic OCP (see, e.g., \cite{21}). In this sense the ideal TCP behaves like an ionic
OCP with $\Gamma=\Gamma_{ii}$. The second moments calculated from Eq.~(21) using
a HNC radial distribution function $g_{ei}(r)$ are shown in Figs.~7 and 8 as a
function of $\Gamma_{ee}$ for hydrogen and aluminum TCPs, respectively. The
dashed straight lines represent $\left<E^{2}\right>_{0}$ for the ideal system. The
other curves
are calculated for different $\delta$ up to the critical values $\sigma_{c}=\Gamma%
_{ei}^{(c)}/{\bar\delta}=\left(\Gamma_{ee}^{(c)}/{\bar\delta}\right)\left(Z^{2}(Z+1)%
\right)^{1/3}$ introduced above. For small $\Gamma_{ee}$ the deviations of the second
moment from $\left<E^{2}\right>_{0}$ are small and the second moment decreases
approximately as $1/\Gamma_{ee}$. But unlike $\left<E^{2}\right>$ of an ionic OCP,
it increases again with the coupling parameter $\Gamma_{ee}$ due to the strong
attractive ion-electron interactions.

The normalized MFDs from PMFEX and MD are compared in Figs.~9-18 where the electric
microfields are scaled in units of the Holtsmark field $E_{H}$ (see Eq.~(25)). For
each distribution we have also calculated the second moment as a control parameter
and found a quite good agreement between Eq.~(21) and the MD simulations. The MFDs
for hydrogen with coupling parameters $\Gamma_{ee}=\Gamma_{ii}=1$ and for Al$^{13+}$
plasmas with $\Gamma_{ee}=0.1$ and $\Gamma_{ii}=7.19$ are shown in Fig.~9 and 10,
respectively. The dashed curves are the Holtsmark MFDs for a TCP with regularized
Coulomb potential. Note that the Holtsmark MFD is $Z$-dependent here (see Eqs.~(13)
and (24)). To demonstrate the importance of attractive interactions we also plotted
the MFDs $P_{0}(E)$ resulting from the corresponding electronic and ionic OCPs with
$\Gamma_{ee}$ and $\Gamma_{ii}$, respectively (open circles). To that end the distribution
$Q_{0}({\bf E})$ of the total field ${\bf E}={\bf E}_{1}+{\bf E}_{2}$ is calculated as
\begin{equation}
Q_{0}\left({\bf E}\right)=\int Q_{e}\left({\bf E}_{1}\right)Q_{i}\left({\bf E}_{2}
\right)\delta\left({\bf E}-{\bf E}_{1}-{\bf E}_{2}\right)d{\bf E}_{1}d{\bf E}_{2}
\equiv \frac{P_{0}(E)}{4\pi E^{2}}
\end{equation}
from the MFD of the ionic OCP at a charged point $Q_{i}\left({\bf E}_{2}\right)$
and of the electronic OCP at a neutral reference point $Q_{e}\left({\bf E}_{1}\right)$.
The distribution $Q_{0}\left({\bf E}\right)$ thus represents the MFD in a TCP
assuming that the ion-electron attractive interaction is switched off.
Here $Q_{e}\left({\bf E}_{1}\right)$ and $Q_{i}\left({\bf E}_{2}\right)$ are taken from
MD simulations of an OCP.

Systematic dependencies of the MFD on $\delta$ and $\Gamma$ are shown in Figs.~11-16.
For fixed $\Gamma$ the maximum of $P(E)$ shifts only slightly to lower field
strengths $E$ with increasing $\delta$, see Figs.~11 and 12, while the maximum itself
increases with $\delta$. This is related to the largest possible single-particle field
$\left|E_{e}(0)\right|=e_{F}/2\delta^{2}$, which an electron can produce at
the ion. Thus the nearest neighbor electronic MFD vanishes for electric fields larger
than $\left|E_{e}(0)\right|$, and smaller $\delta$ will result in larger contributions to
$P(E)$ at higher fields $E$ with a corresponding reduction of $P(E)$ at small fields.
In order to demonstrate the enhanced probability of large fields at small $\delta$
and the behavior of PMFEX and MD treatments at large fields, the MFD is plotted in Figs.~13
and 14 in a double-logarithmic manner. From Fig.~13 it can be deduced that the behavior
of the MFD at large fields in H$^{+}$ plasma with $\bar\delta=0.2$ and $\Gamma_{ee}=1$
is similar to the nearest neighbor electronic distribution considered in
detail in Ref.~\cite{43}. In this case the MFD is strongly reduced at $E>\left|E_{e}(0)%
\right|\simeq 12.5E_{H}$.

For fixed $\delta$ and increasing $\Gamma$ the MFDs for hydrogen (Fig.~15) and Al$^{13+}$
(Fig.~16) show different behavior. For hydrogen, like for an ionic OCP, the growing
correlations shift the maximum of the MFD towards lower electric fields. In the
Al$^{13+}$-TCP, $P(E)$ first follows this trend, but then, for further increasing
$\Gamma$, the maximum turns back to higher field strengths. This can be attributed to
the growing contribution of the attractive electron-ion interaction and close ion-electron
configurations, which are particularly important for a TCP with highly charged ions.
These dependencies are very well reproduced by the PMFEX predictions.

The agreement with the MD data is nearly perfect in most of the studied cases,
both for the H$^{+}$-TCP and the Al$^{13+}$-TCP (Figs.~9-16). The PMFEX approximation
remains accurate also up to high electric fields where the MD data are characterized
by strong fluctuations (see Figs.~13 and 14). Deviations emerge only for strongly
coupling situations with large $\Gamma$ and $\sigma$. One example is the case of
strongly coupled hydrogen with $\Gamma_{ee}=1$ and $\bar\delta=0.2$, i.e. $\sigma\simeq
6.3$ (dotted line and open triangles in Fig.~15, see also Fig.~13).
Here PMFEX and MD results differ considerably, although the HNC treatment is accurate
in this case (see Fig.~2).~To understand this feature better we recall that within
PMFEX the Fourier transformed MFD, $T(K)$, fulfills exactly the second moment
relation~(21) in the limit $K\rightarrow 0$. As discussed above only electrons contribute
to the second moment since the role of ions is negligible (the second term in Eq.~(21)).
Because of Eq.~(36) small values of $K$ correspond to large values of the local electric
field. One expects therefore that the PMFEX yields good results if there are many
electrons near the ion. On the other hand, for a large electron-electron repulsion
$\Gamma_{ee}=1$ and a light ion like hydrogen, the electrons tend to exert only
small fields in the ion, for which the quality of the PMFEX is less obvious.

With increased coupling also the shape of the
MFD starts to change. First by a broadening of the maximum, and then by the
appearance of a shoulder as also reported in Ref.~\cite{43} which then gets more
and more pronounced and finally develops into a second maximum. For the Al$^{13+}$ TCP
in a parameter regime still below the critical values, the MFD is characterized
here by the formation of the characteristic shape shown in Fig.~17. The HNC
approximation still gives the correct $g_{ii}(r)$ and $g_{ei}(r)$ and the PMFEX
well reproduces the broadening and the specific shape of the MFD. With respect
to $g_{ee}(r)$, however, deviations between HNC and MD emerge (similar to those shown
for $\delta=0.4a$, $\Gamma_{ee}=0.2$ in Fig.~6), although the electron-electron coupling
($\Gamma_{ee}=0.1$) is still small. The strong ion-electron coupling increases the
electron density near the ion which introduces additional correlations between electrons,
see the discussion above in Sec.~VB. This will, however, not affect the quality of the
PMFEX approximation, since the $g_{ee}(r)$ is not needed for the calculation of the MFD
at the impurity ion (see Eqs.~(13) and (36)). An example for a second maximum is given
by the strongly coupled hydrogen of Fig.~18, where the parameters are close to the
critical values. This regime is characterized by the population of bound states
and the formation of a separate contribution to $P(E)$ at high fields which is
mainly due to the electrons. Here occur significant deviations between the HNC approach
and MD simulations in the RDF $g_{ei}(r)$ (see Fig.~3) and the PMFEX approximation cannot predict the shape of the MFD, even not qualitatively. But, for coupling parameters,
where a classical approach is justified, i.e. when bound states are unimportant, the
PMFEX approach turns out to be a very reliable method for calculating the MFD of a TCP
with attractive interaction.

\section{Discussion and Conclusion}

In this paper our objective was to investigate the microfield distributions in a
two-component plasmas with attractive electron-ion interactions. Attention has been
focused on testing the predictions of the PMFEX approximation based on the HNC
treatment of static correlations by confronting it with the MFDs obtained from
MD simulations. One of the basic assumptions of the model considered here is the
regularization of the attractive Coulomb interaction at short distances to introduce
quantum diffraction effects in the employed classical approach.

Two specific rather distinct cases, H$^{+}$ ($n_{e}=n_{i}$) and Al$^{13+}$
($n_{e}=13n_{i}$) two-component plasmas with symmetric and largely asymmetric
density distributions between plasma species were considered. For
simplicity we assume bare Coulomb electron-electron and ion-ion interactions while
the parameter $\bar\delta$ for the regularized ion-electron potential varies from
$0.1$ to $0.4$. The coupling strength between plasma particles is measured by the
coupling parameters $\Gamma_{\alpha\beta}$ with $\alpha,\beta=e,i$ and by the
ion-electron potential at the origin in units of $k_{B}T$, $\sigma=\Gamma_{ei}%
/{\bar\delta}$. Our treatment is limited to a parameter regime with $\sigma<\sigma_{c}%
(Z,{\bar\delta})$, where the critical value $\sigma_{c}$ for $0.1\leq {\bar\delta}\leq 0.4$
varies $8.32\leq \sigma_{c}\leq 13.4$ and $6.66\leq \sigma_{c}\leq 7.33$ for H$^{+}$ and
Al$^{13+}$ TCPs, respectively. Within this parameter regime the $g_{\alpha\beta}(r)$
from the HNC equations agree well with the MD simulations. Beyond these critical $\sigma$
the HNC equations do either not converge or end up in unphysical solutions while the
MD simulations remain effective at these strong coupling regimes. A further increase
of the coupling parameters also leads to the formation of classical strongly bound
electronic states with no corresponding quantum counterpart. Also the microfield
distributions obtained from the HNC via the PMFEX approximation agree excellently with
the MFDs from the MD simulations except of some cases close to the critical $\Gamma$,
$\sigma$. This is somewhat surprising since a similar approximation studied for the
OCP deviates from MD simulations (see, e.g., \cite{21,22}). Therefore we have also
tested the PMFEX approximation for an OCP, which in contrast to the TCP turns out
to be poor when compared with MD simulations although the exact second moment is
satisfied within the PMFEX. The success of the PMFEX approximation for the TCP is
a consequence of the attractive interaction and is related to the additional positive
electronic part in Eq.~(36) which accounts for the electric fields created by the
electrons at the ions. Obviously, the attractive interactions in a TCP favour
configurations with large electric fields created at the ion which are well described
within the PMFEX approximation. On the other hand, in a regime dominated by small
local fields and hence by small local electronic density the PMFEX deviates from the
MD. This feature has been clearly observed for a single ion embedded in an electronic
OCP in Ref.~\cite{43}. For the TCP, an example is the case of strongly coupled hydrogen
with $\Gamma_{ee}=1$, $\delta=0.2a$ in Figs.~13 and 15. Here some improvement of the
PMFEX scheme is required. Such work and the application of PMFEX to the case of a
neutral radiator are in progress.

\begin{acknowledgments}
This work was supported by the Bundesministerium f\"ur Bildung und Forschung
(BMBF) under contract no 06ER128.
\end{acknowledgments}

\appendix

\section{The effective electric fields}

In order to reduce the three-dimensional integration in Eq.~(34) to a one-dimensional
integration and to express the effective fields through scalar potentials $u_{\alpha
R}(r)$ we consider the following expression
\begin{equation}
\int d\mathbf{r}_{1}u_{\beta R}\left(r_{1}\right)\left[g_{\alpha\beta}\left(\left|%
\mathbf{r}-\mathbf{r}_{1}\right|\right)-1\right]=4\pi\int_{0}^{\infty}G^{(0)}_{\beta}%
(r,\rho)\left[g_{\alpha\beta}(\rho)-1\right] \rho d\rho,
\end{equation}
where
\begin{equation}
G^{(0)}_{\beta}(r,\rho)=\frac{1}{2r}\int_{\left|r-\rho\right|}^{r+\rho} u_{\beta R}%
(r^{\prime}) r^{\prime} dr^{\prime}.
\end{equation}
Obviously, the gradient of Eq.~(A1) yields the second term in Eq.~(34). Consequently,
recalling the spherical symmetry of the single particle fields, Eq.~(34)
can be alternatively expressed through one-dimensional integrals
\begin{equation}
{\cal E}_{\alpha}\left(r\right)=E_{\alpha}\left(r\right)+\frac{4\pi e_{F}}
{g_{\alpha R}\left(r\right)}\sum_{\beta}q_{\beta}n_{\beta}\int_{0}^{\infty} %
G^{(1)}_{\beta}\left(r,\rho\right)\left[g_{\alpha\beta}\left(\rho\right)-1\right]
\rho d\rho.
\end{equation}
Here
\begin{equation}
G^{(1)}_{\beta}\left(r,\rho\right)=-\frac{\partial }{\partial r}
G^{(0)}_{\beta}\left(r,\rho\right).
\end{equation}
For the regularized Coulomb interaction the last expression yields for $r<\rho$
and $r>\rho$, respectively
\begin{equation}
G^{(1)}_{\beta}\left(r,\rho\right)=\frac{\delta_{\beta R}}{r^{2}}e^{-\rho/\delta_
{\beta R}}\left[\frac{r}{\delta_{\beta R}} {\rm ch}\left(\frac{r}{\delta_{\beta R}}
\right)-{\rm sh} \left(\frac{r}{\delta_{\beta R}}\right)\right],
\end{equation}
\begin{equation}
G^{(1)}_{\beta}\left(r,\rho\right)=\frac{\delta_{\beta R}}{r^{2}} \left[\frac{\rho%
}{\delta_{\beta R}}-{\rm sh}\left(\frac{\rho}{\delta_{\beta R}}%
\right)\left(1+\frac{r}{\delta_{\beta R}}\right)e^{-r/\delta_{\beta R}}\right].
\end{equation}
For bare Coulomb interaction Eqs.~(A5) and (A6) for $r<\rho$ and $r>\rho$ are
reduced to $G^{(1)}_{\beta}\left(r,\rho\right)=0$ and  $G^{(1)}_{\beta}\left
(r,\rho\right)=\rho/r^{2}$, respectively.

\newpage

\begin{figure}
\includegraphics[width=8cm]{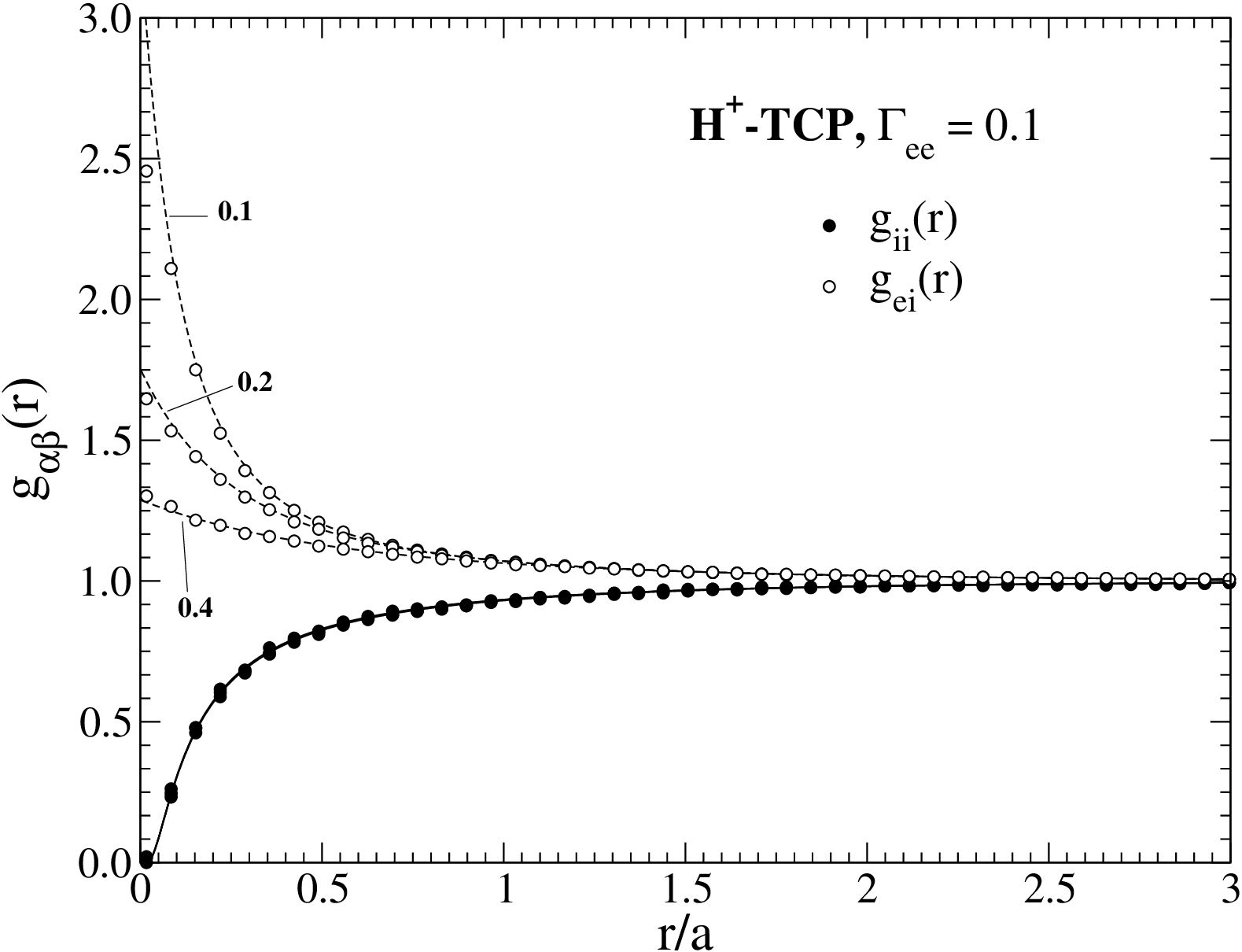}
\caption{RDFs $g_{\alpha\beta}(r)$ for a H$^{+}$ plasma with fixed $\Gamma_{ee}=0.1$
and ${\bar\delta}=0.1$, $0.2$, and $0.4$. The lines correspond to the HNC
approximation while the symbols denote the MD simulations. The different lines and
symbols represent $g^{\rm HNC}_{ii}\equiv g^{\rm HNC}_{ee}$ (solid lines), $g^{\rm HNC}_{ei}$ (dashed lines), $g^{\rm MD}_{ii}=g^{\rm MD}_{ee}$ (filled circles), $g^{\rm
MD}_{ei}$ (open circles). The numbers indicate the values of ${\bar\delta}$.}
\end{figure}

\begin{figure}
\includegraphics[width=8cm]{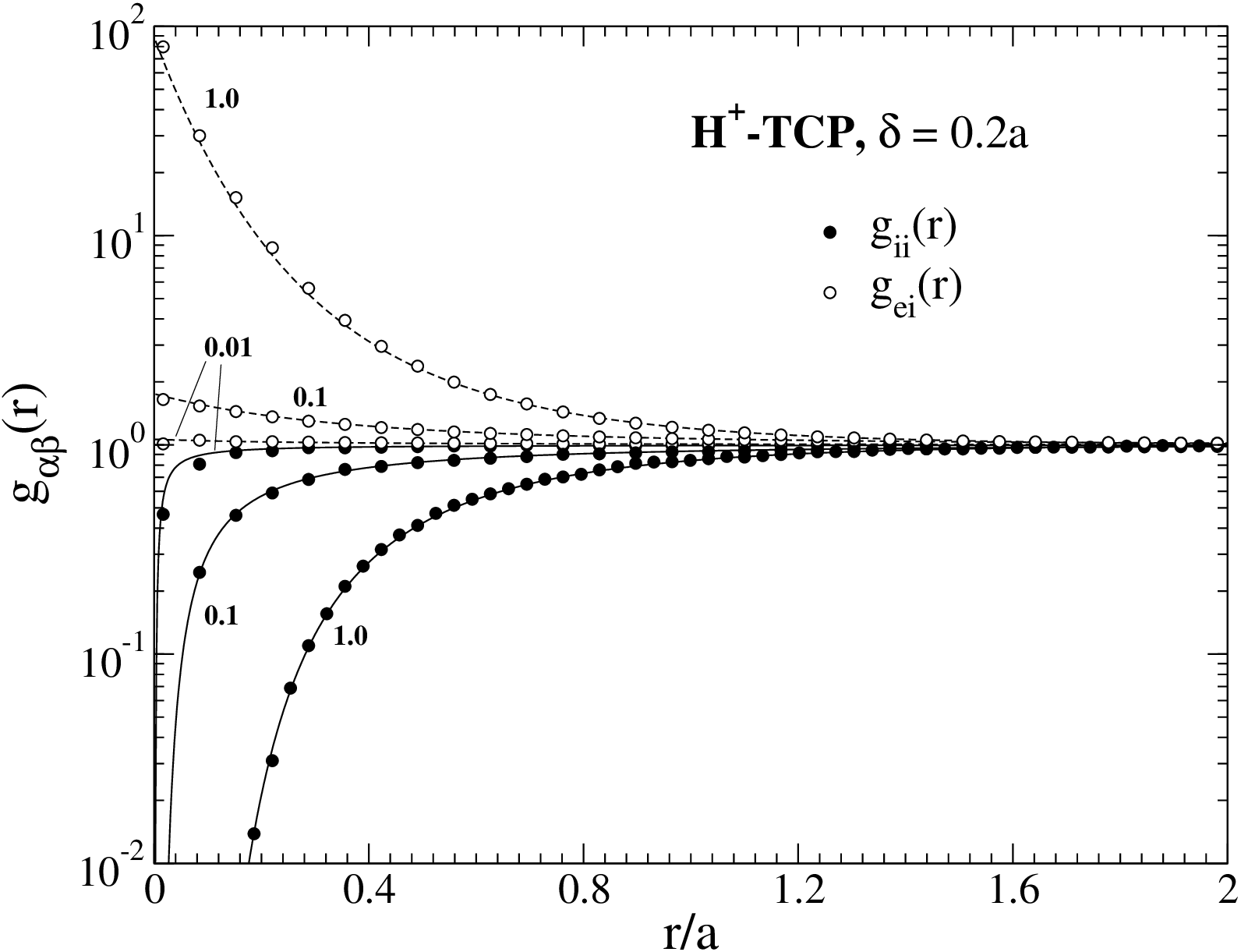}
\caption{Same as Fig.~1 with fixed ${\bar\delta}=0.2$ and $\Gamma_{ee}=0.01$, $0.1$,
and $1.0$. Here the numbers indicate the values of $\Gamma_{ee}$.}
\end{figure}

\begin{figure}
\includegraphics[width=8cm]{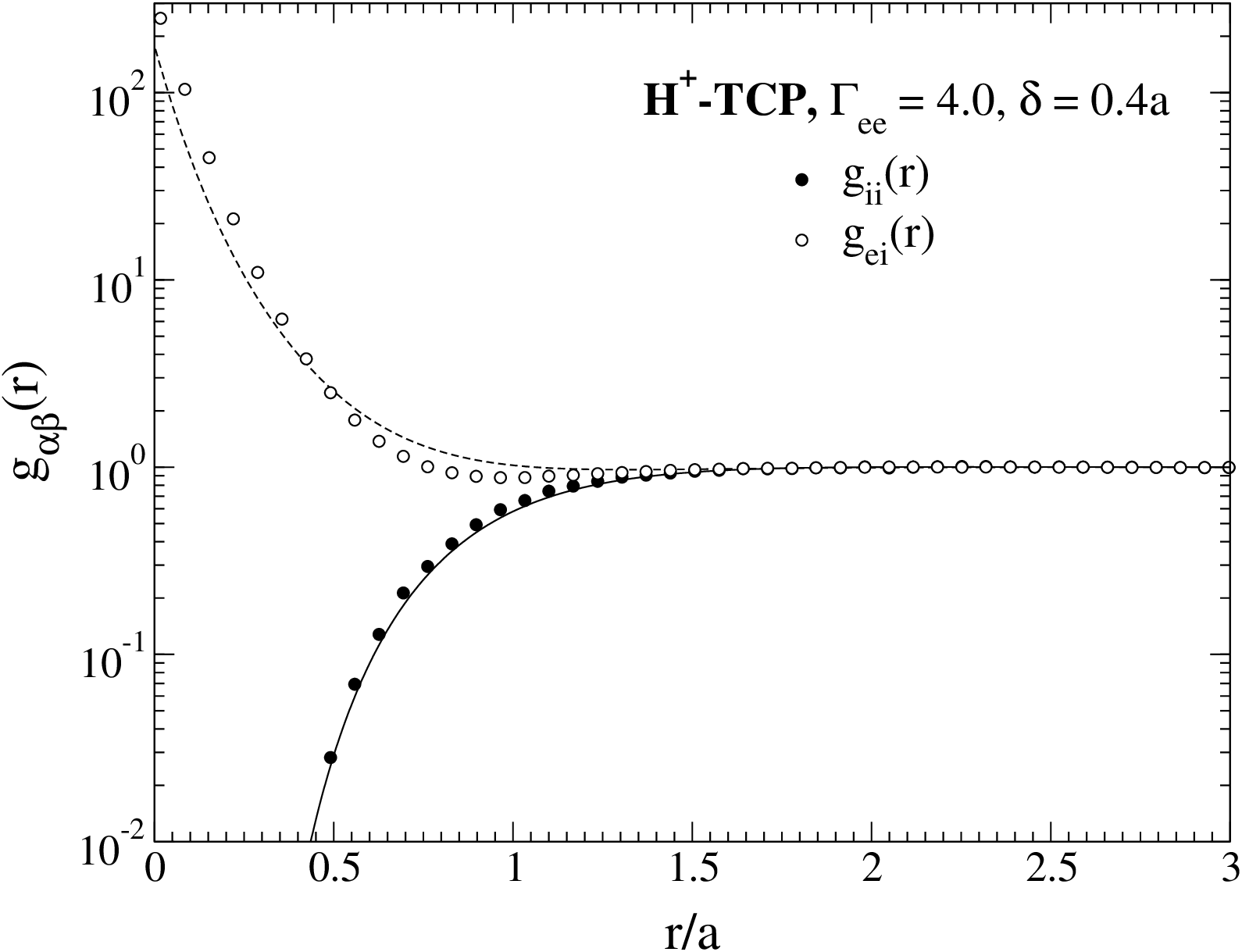}
\caption{Same as Fig.~1 for $\Gamma_{ee}=4.0$ and ${\bar\delta}=0.4$.}
\end{figure}

\begin{figure}
\includegraphics[width=8cm]{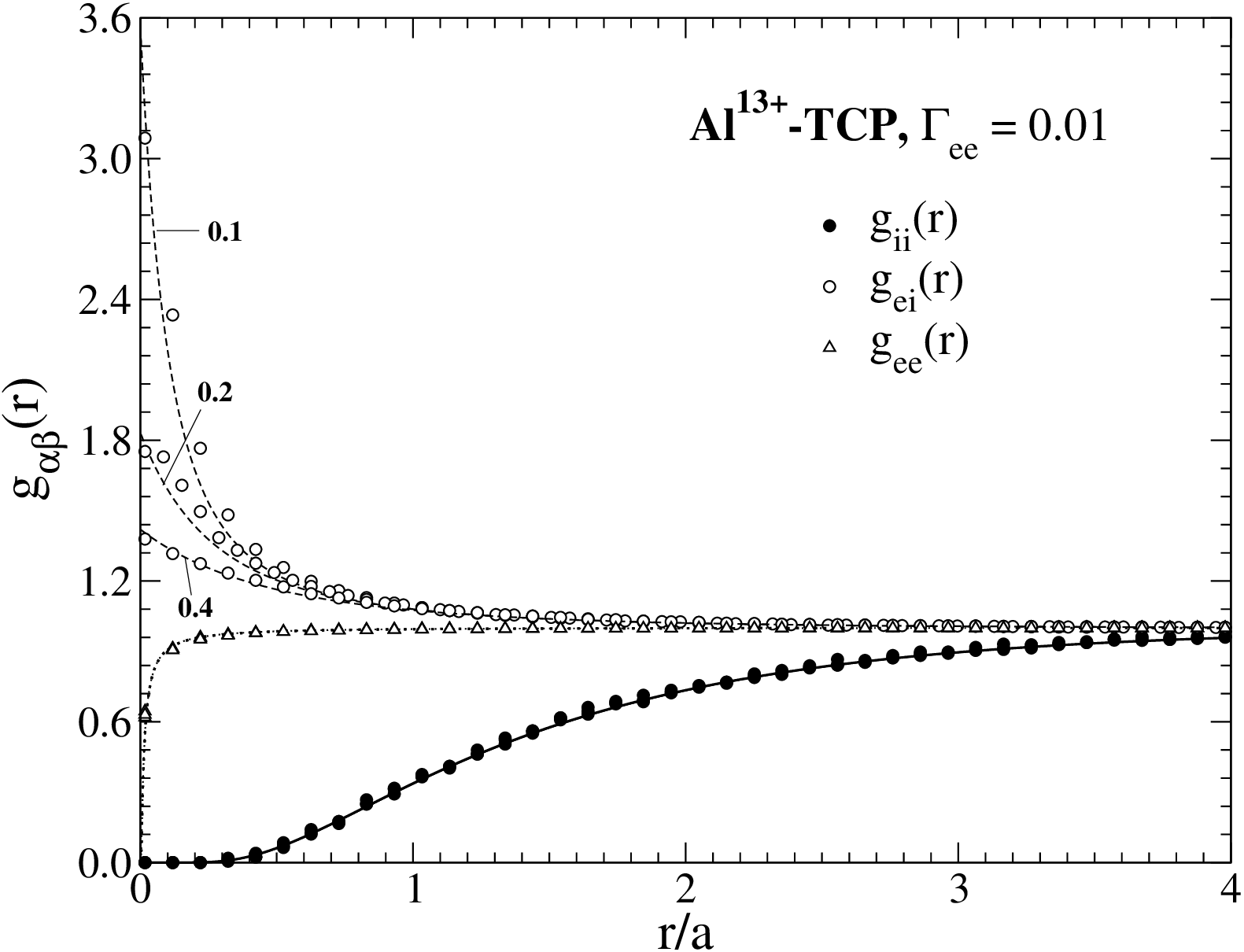}
\caption{Same as Fig.~1 for a Al$^{13+}$ plasma with fixed $\Gamma_{ee}=0.01$ and
${\bar\delta}=0.1$, $0.2$, and $0.4$. The dotted lines and the triangles represent
$g^{\rm HNC}_{ee}$ and $g^{\rm MD}_{ee}$, respectively.}
\end{figure}

\begin{figure}
\includegraphics[width=8cm]{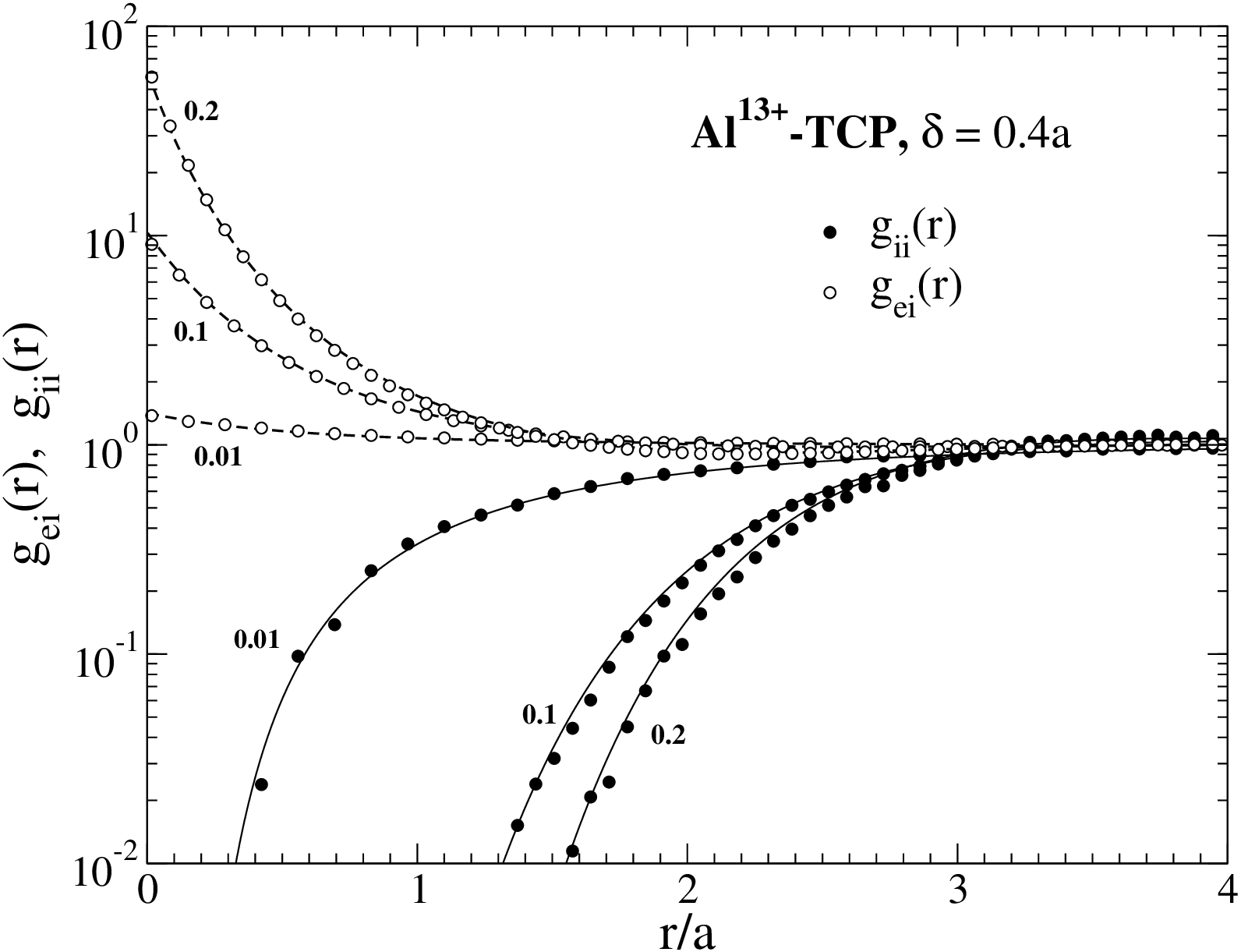}
\caption{Same as Fig.~4 for the RDFs $g_{ei}(r)$ and $g_{ii}(r)$ with ${\bar\delta}=0.4$,
and $\Gamma_{ee}=0.01$, $0.1$, and $0.2$ as indicated by the numbers.}
\end{figure}

\begin{figure}
\includegraphics[width=8cm]{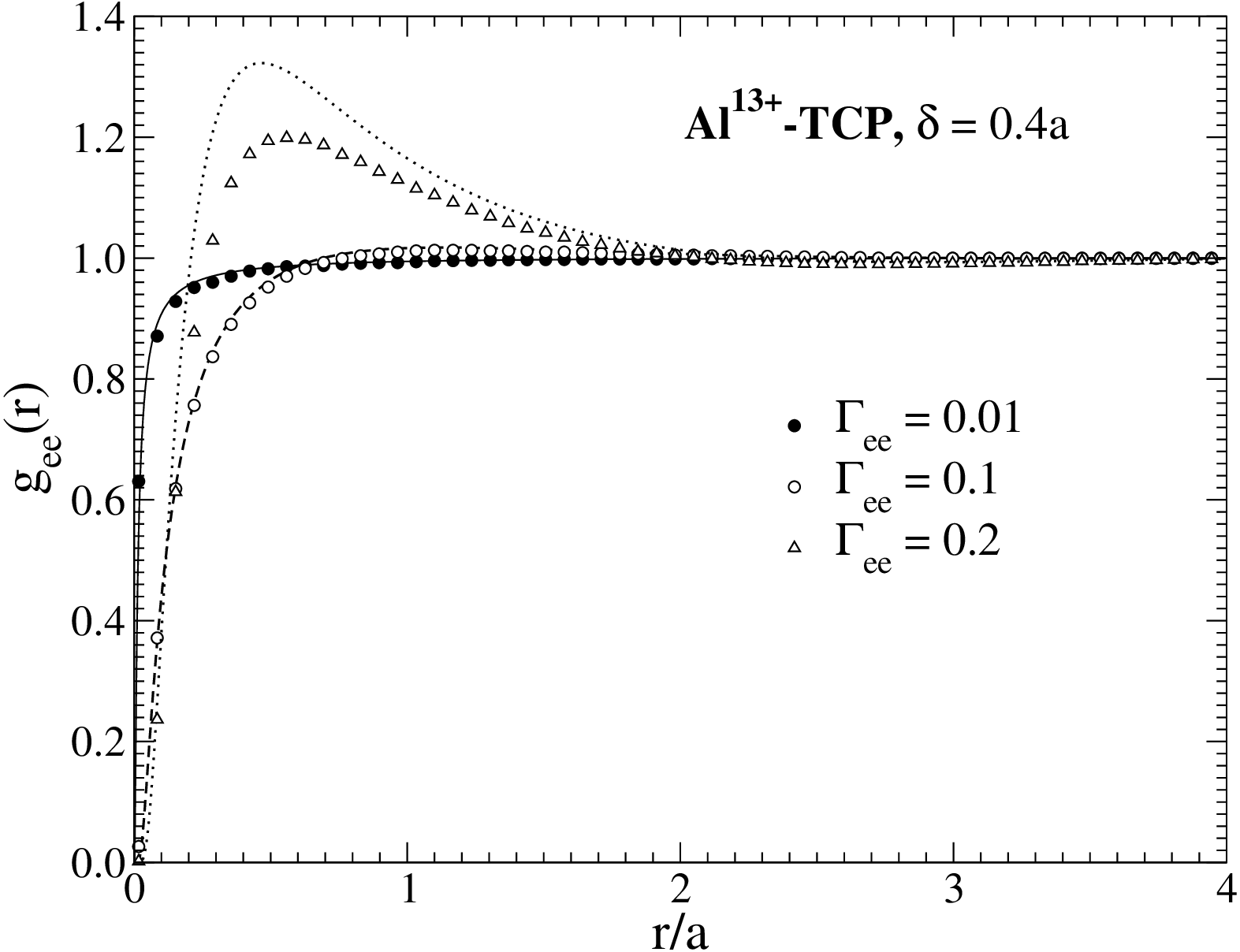}
\caption{Same as Fig.~5 for $g_{ee}(r)$.}
\end{figure}

\begin{figure}
\includegraphics[width=8cm]{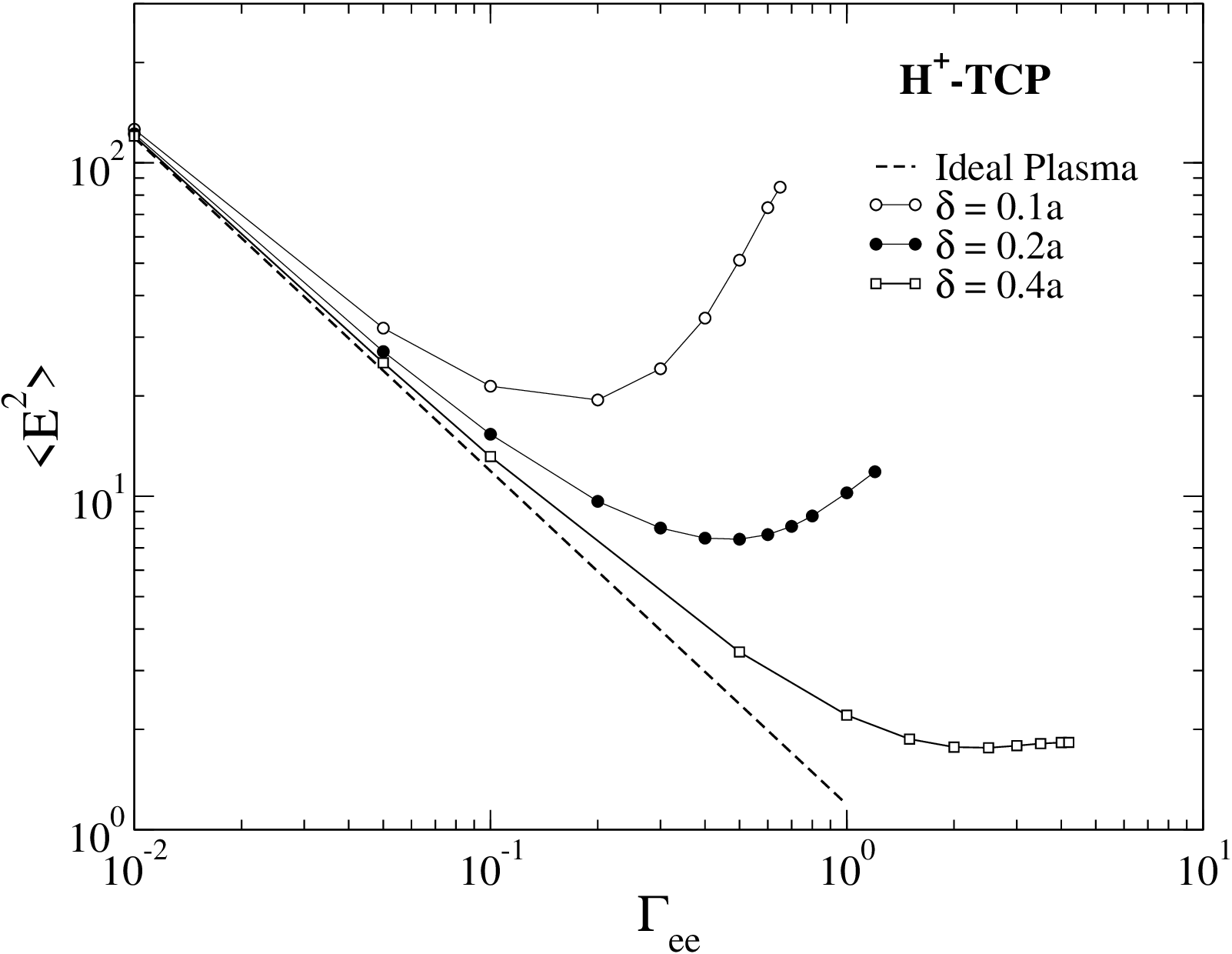}
\caption{The second moment $\left<E^{2}\right>$ of the MFD (in units $E_{H}^{2}$, see
Eq.~(25)) obtained from the HNC scheme using Eq.~(21) as a function of $\Gamma_{ee}$
for a H$^{+}$-TCP. The dashed line corresponds to the limiting case of an ideal plasma
(see the text for details). The lines with open and filled circles and squares represent
the second moments for ${\bar\delta}=0.1$, ${\bar\delta}=0.2$ and ${\bar\delta}=0.4$,
respectively.}
\end{figure}

\begin{figure}
\includegraphics[width=8cm]{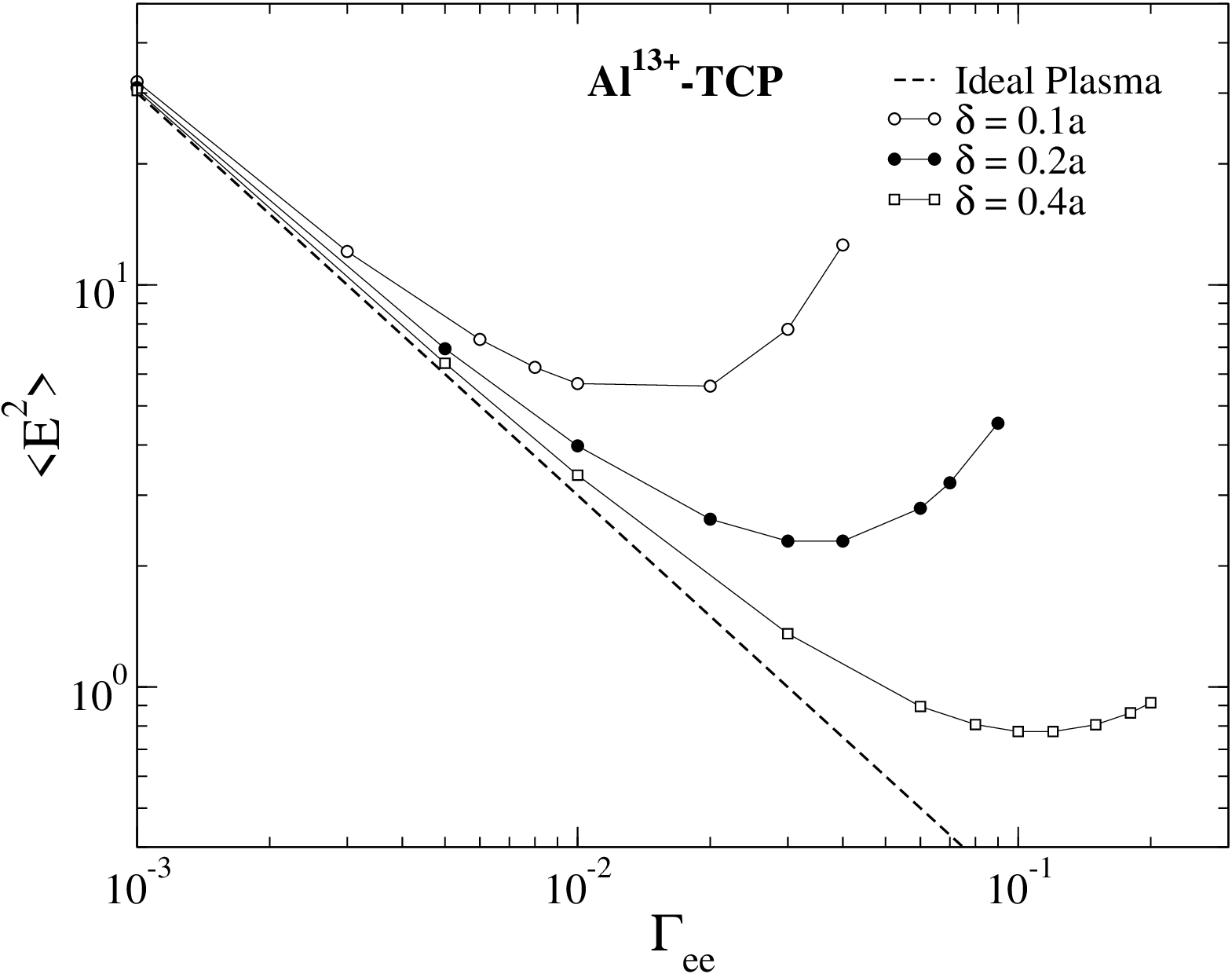}
\caption{Same as Fig.~7 for a Al$^{13+}$-TCP.}
\end{figure}

\begin{figure}
\includegraphics[width=8cm]{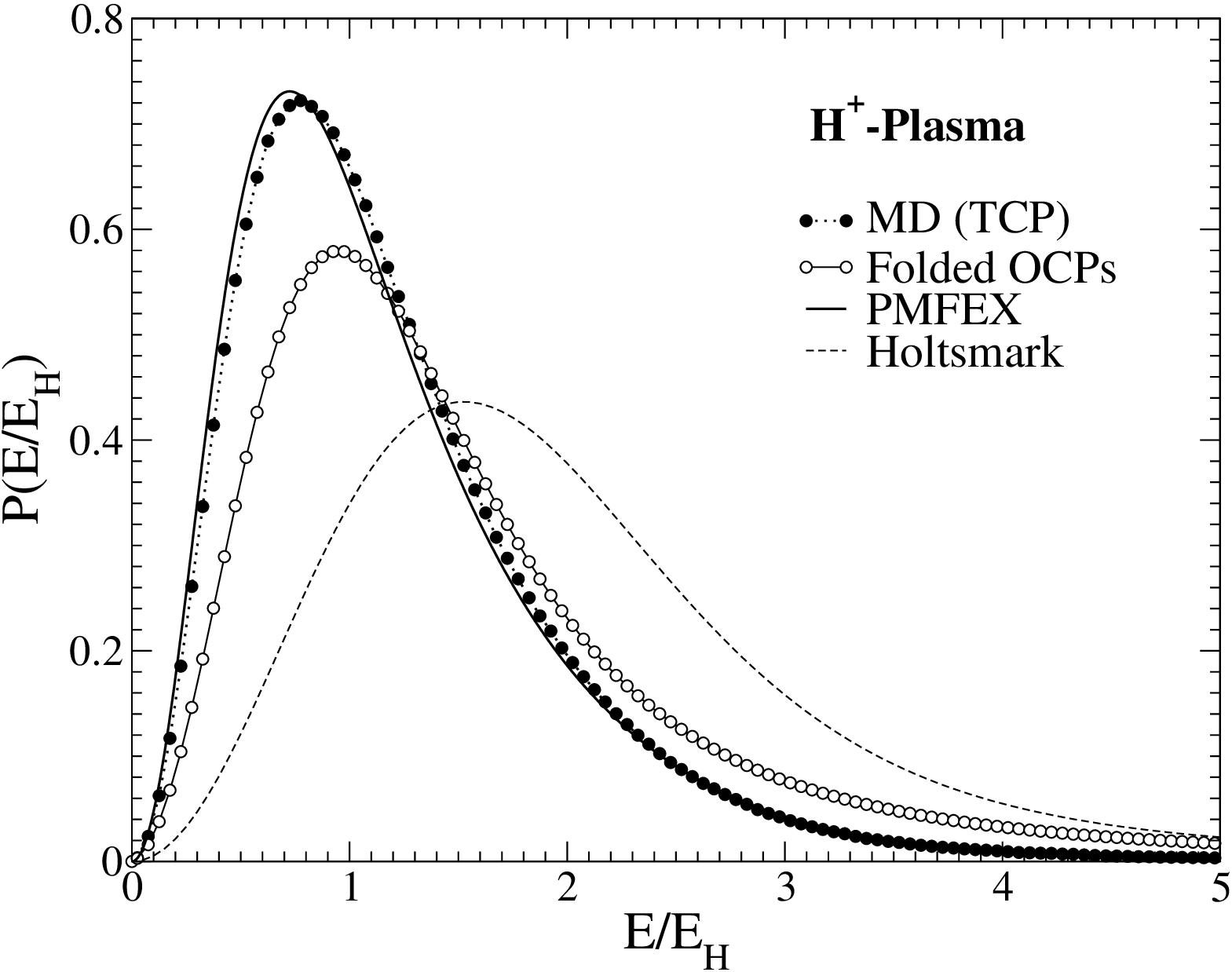}
\caption{Normalized electric microfield distributions for a hydrogen plasma with
$\Gamma_{ee}=\Gamma_{ii}=1$ and $\bar\delta=0.4$ as a function of the electric
field in units of $E_{H}$, Eq.~(25). The filled circles represent the MFD from the
MD simulations and the solid curve the results of the PMFEX. The open circles are
the MFD obtained from the folding of an electronic and an ionic OCP, see Eq.~(42).
The Holtsmark distribution (see Eqs.~(13) and (24)) is shown as a dashed line.}
\end{figure}

\begin{figure}
\includegraphics[width=8cm]{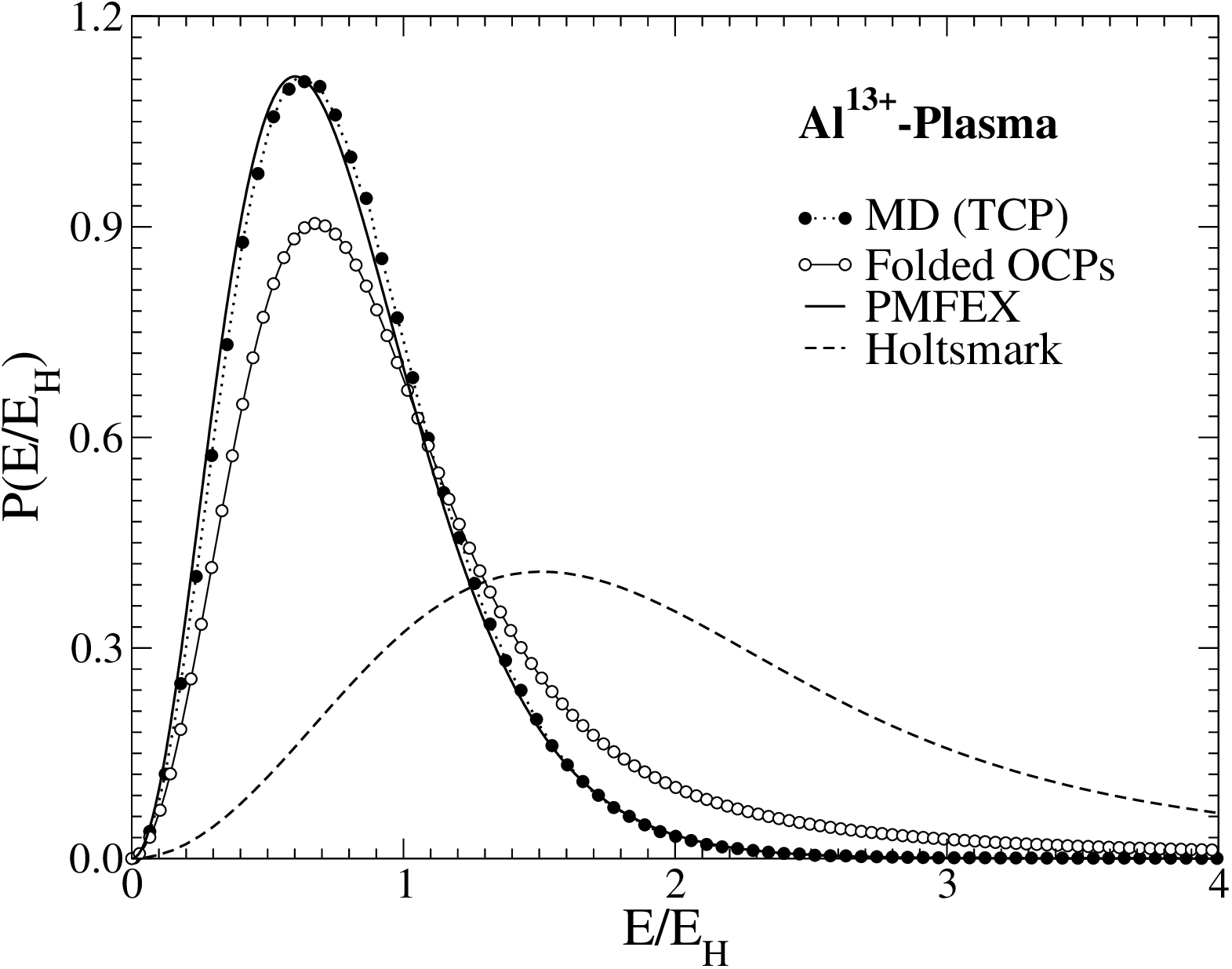}
\caption{Same as Fig.~9 for a Al$^{13+}$ plasma with $\Gamma_{ee}=0.1$ and
$\Gamma_{ii}=7.2$.}
\end{figure}

\begin{figure}
\includegraphics[width=8cm]{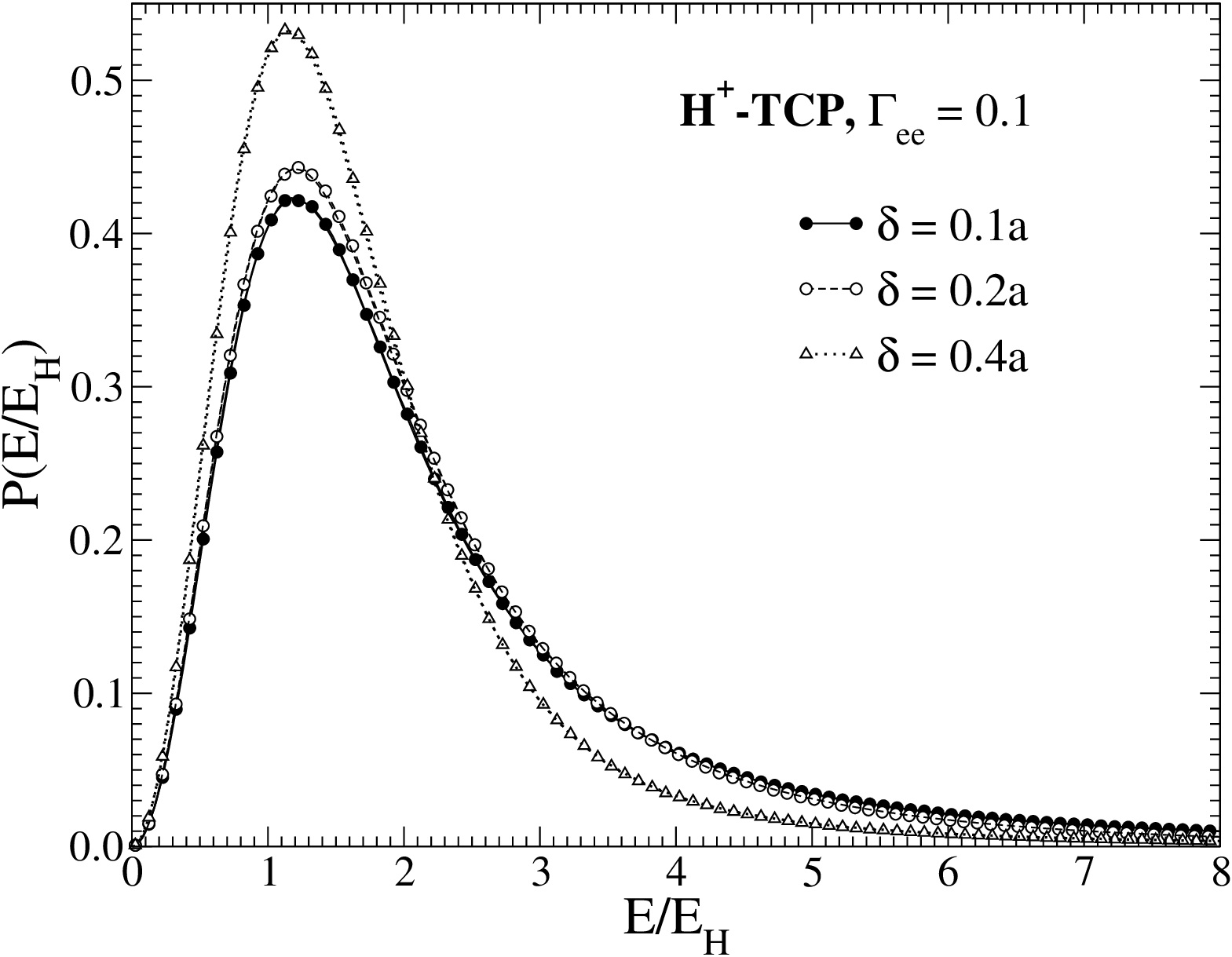}
\caption{Normalized electric microfield distributions for H$^{+}$-plasmas. The lines
with and without symbols correspond to MD simulations and PMFEX approximation,
respectively. $\Gamma_{ee}=0.1$ and ${\bar\delta}=0.1$ (solid lines),~$0.2$
(dashed lines), and~$0.4$ (dotted lines).}
\end{figure}

\begin{figure}
\includegraphics[width=8cm]{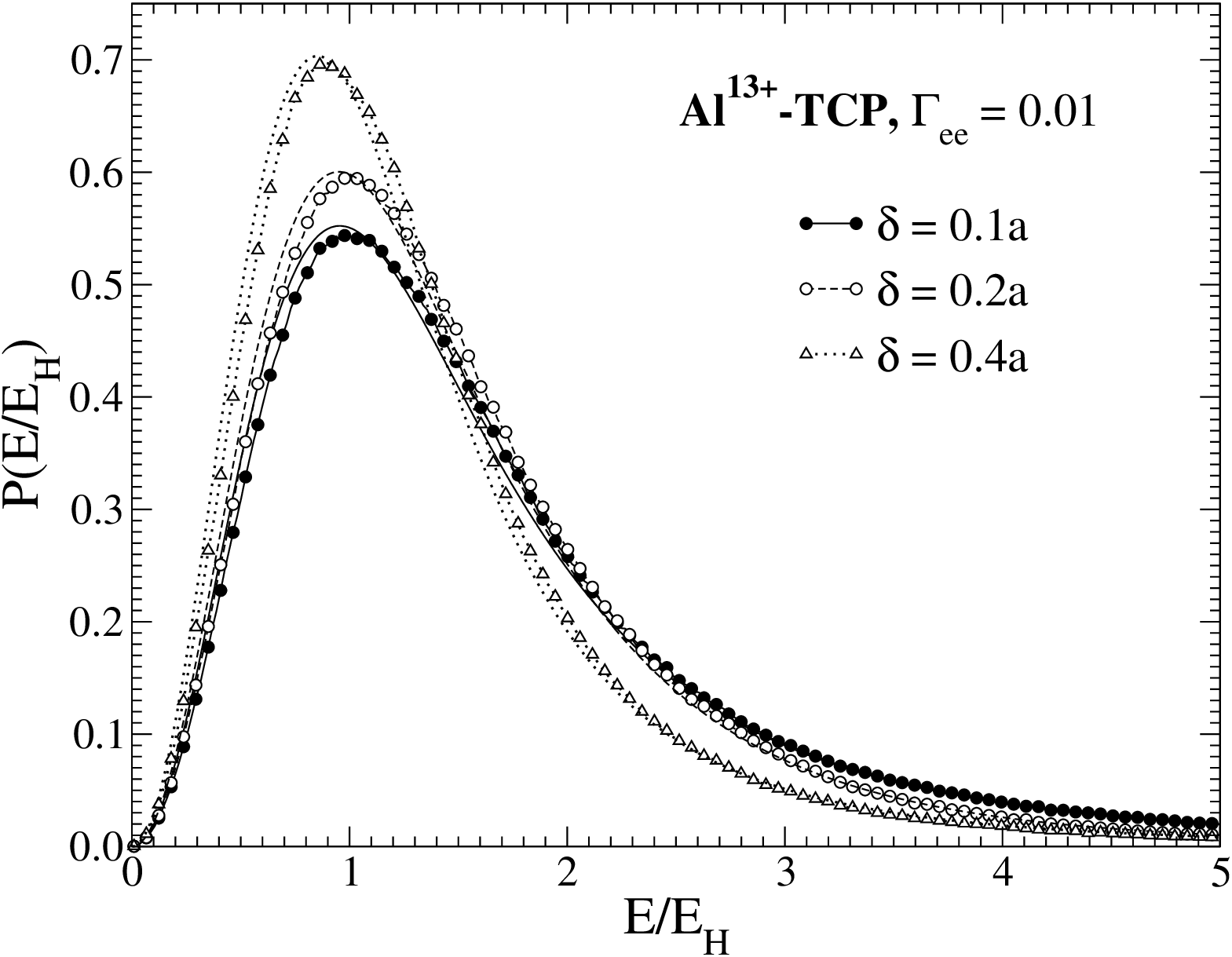}
\caption{Same as Fig.~11 for Al$^{13+}$-plasmas with $\Gamma_{ee}=0.01$.}
\end{figure}

\begin{figure}
\includegraphics[width=8cm]{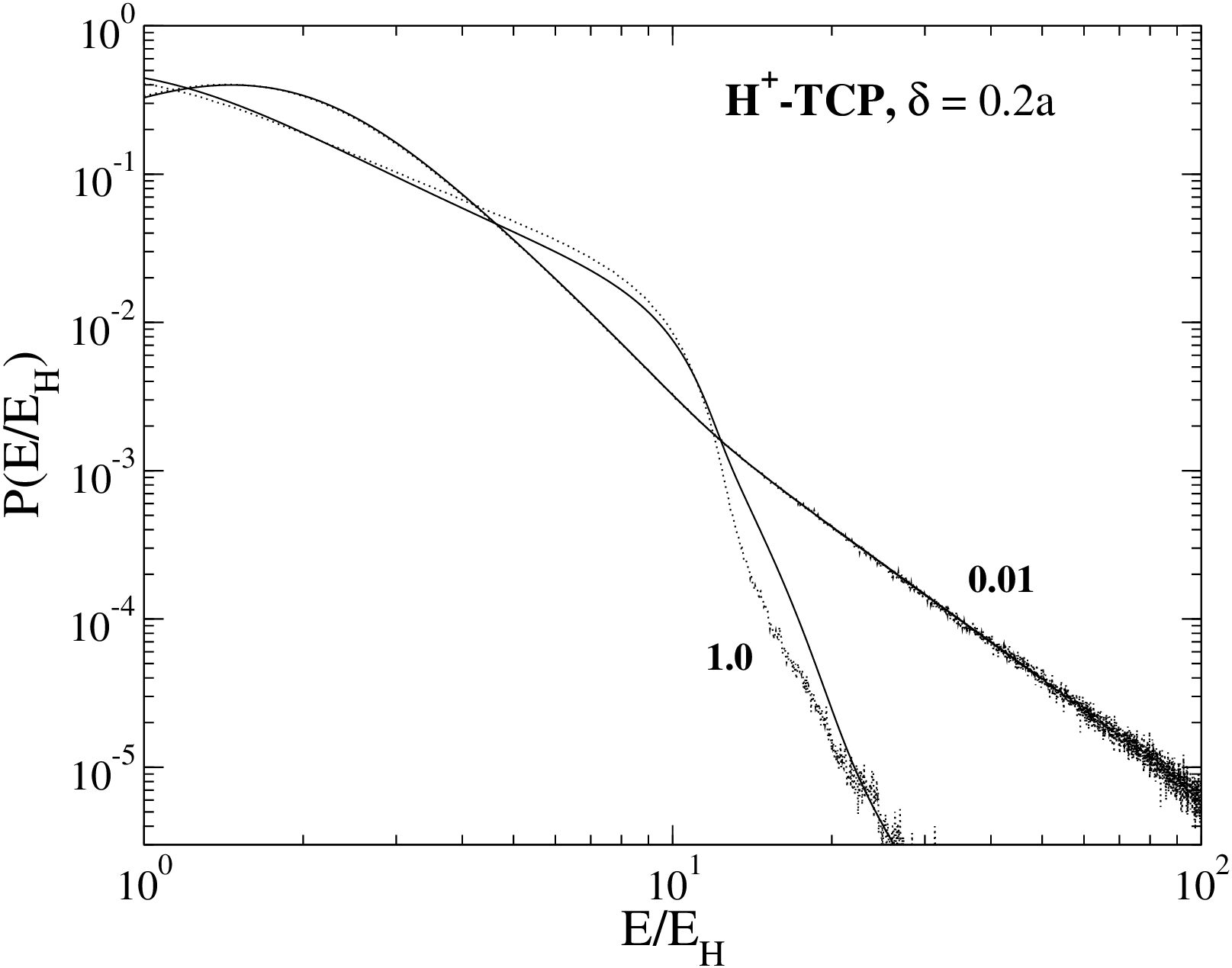}
\caption{MFDs in double-logarithmic plots for H$^{+}$ plasmas with $\bar\delta =0.2$
and $\Gamma_{ee}=0.01$, $\Gamma_{ee}=1.0$ as indicated by the numbers. Here the solid
curves represent the PMFEX approximation and the dotted curves MD simulations, respectively.}
\end{figure}

\begin{figure}
\includegraphics[width=8cm]{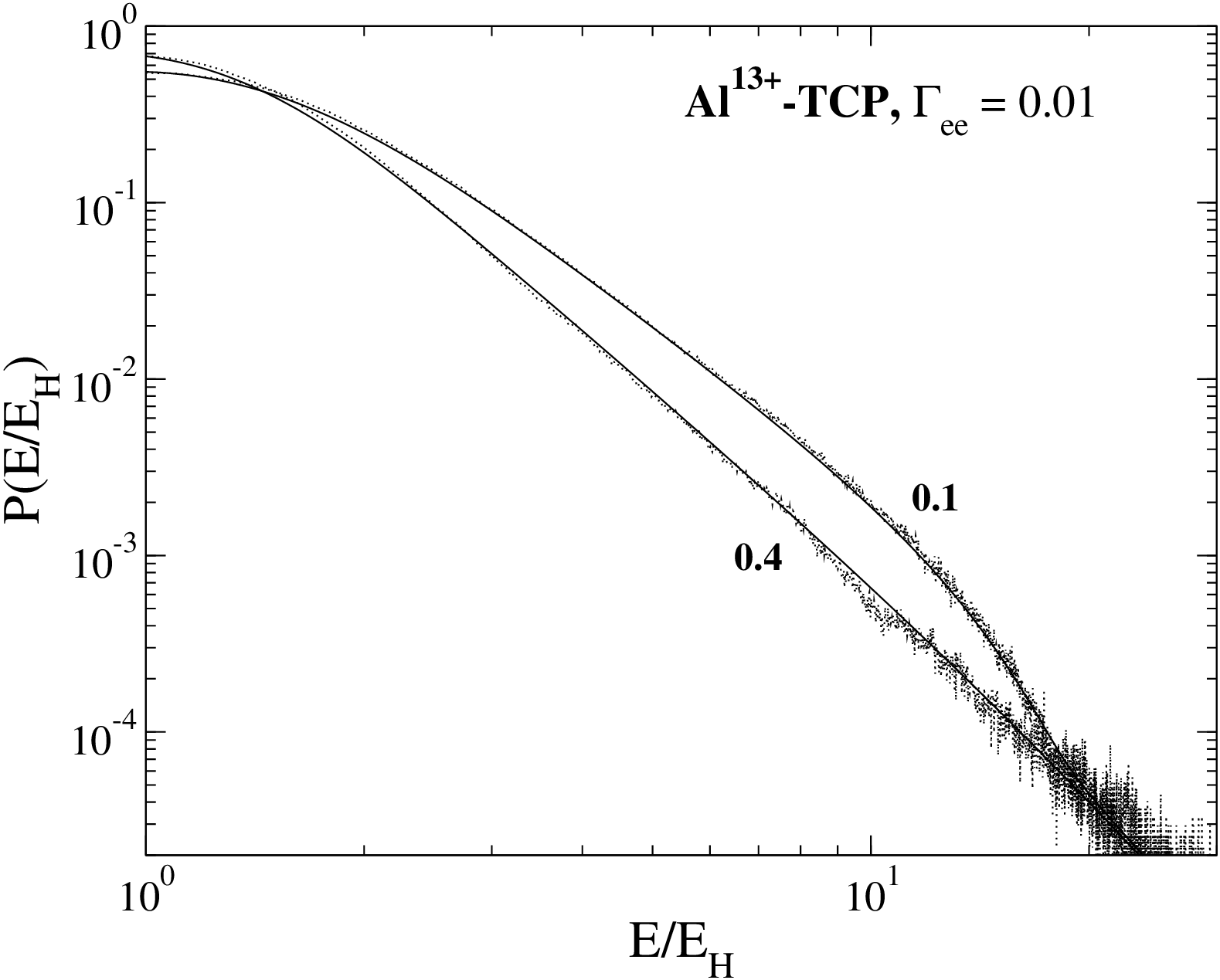}
\caption{Same as Fig.~13 for Al$^{13+}$ plasmas with $\Gamma_{ee}=0.01$ and
$\bar\delta =0.1$, $\bar\delta =0.4$ as indicated by the numbers.}
\end{figure}

\begin{figure}
\includegraphics[width=8cm]{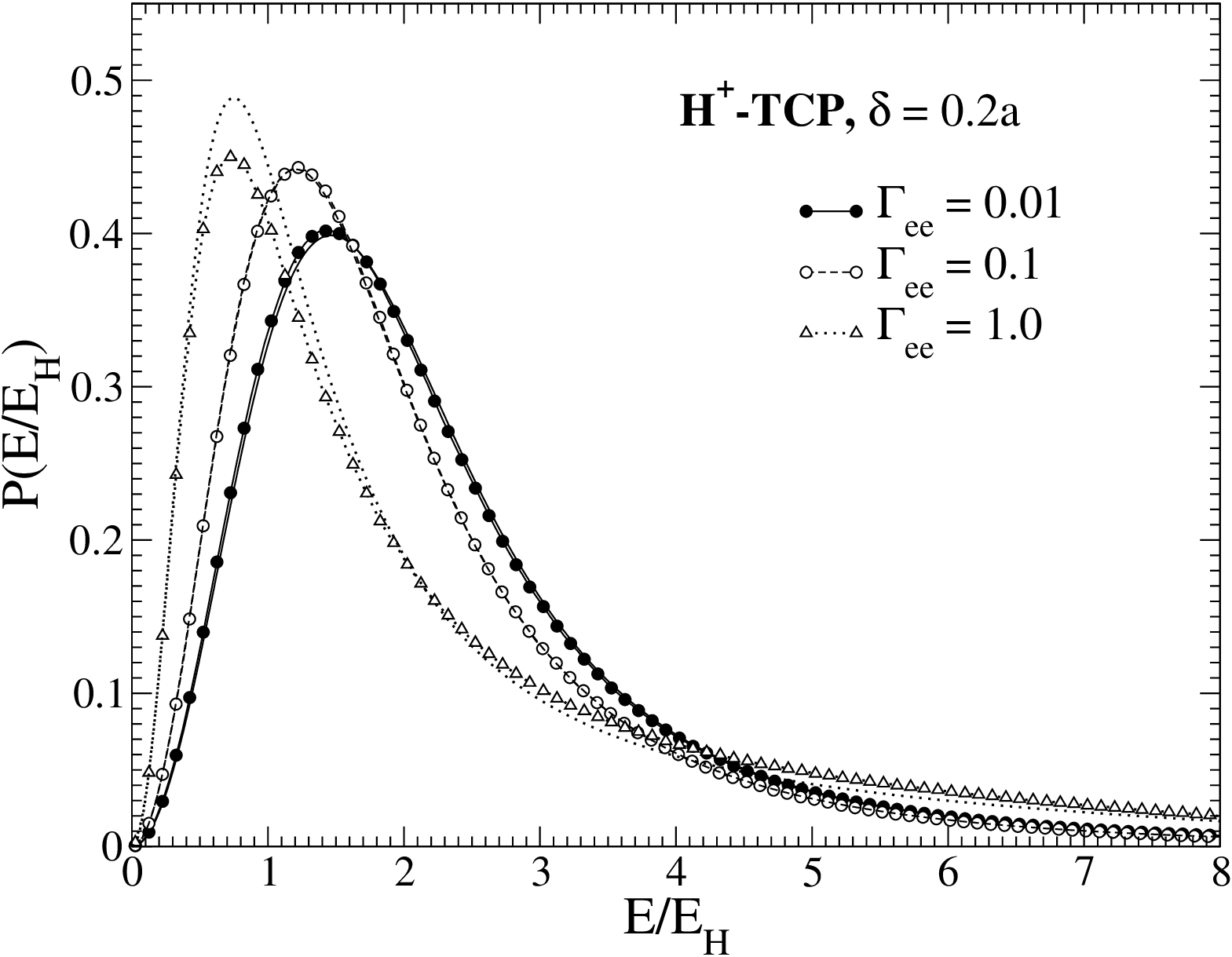}
\caption{Same as Fig.~11 with ${\bar\delta}=0.2$ and $\Gamma_{ee}=0.01$ (solid
lines),~$0.1$ (dashed lines), and~$1.0$ (dotted lines).}
\end{figure}

\begin{figure}
\includegraphics[width=8cm]{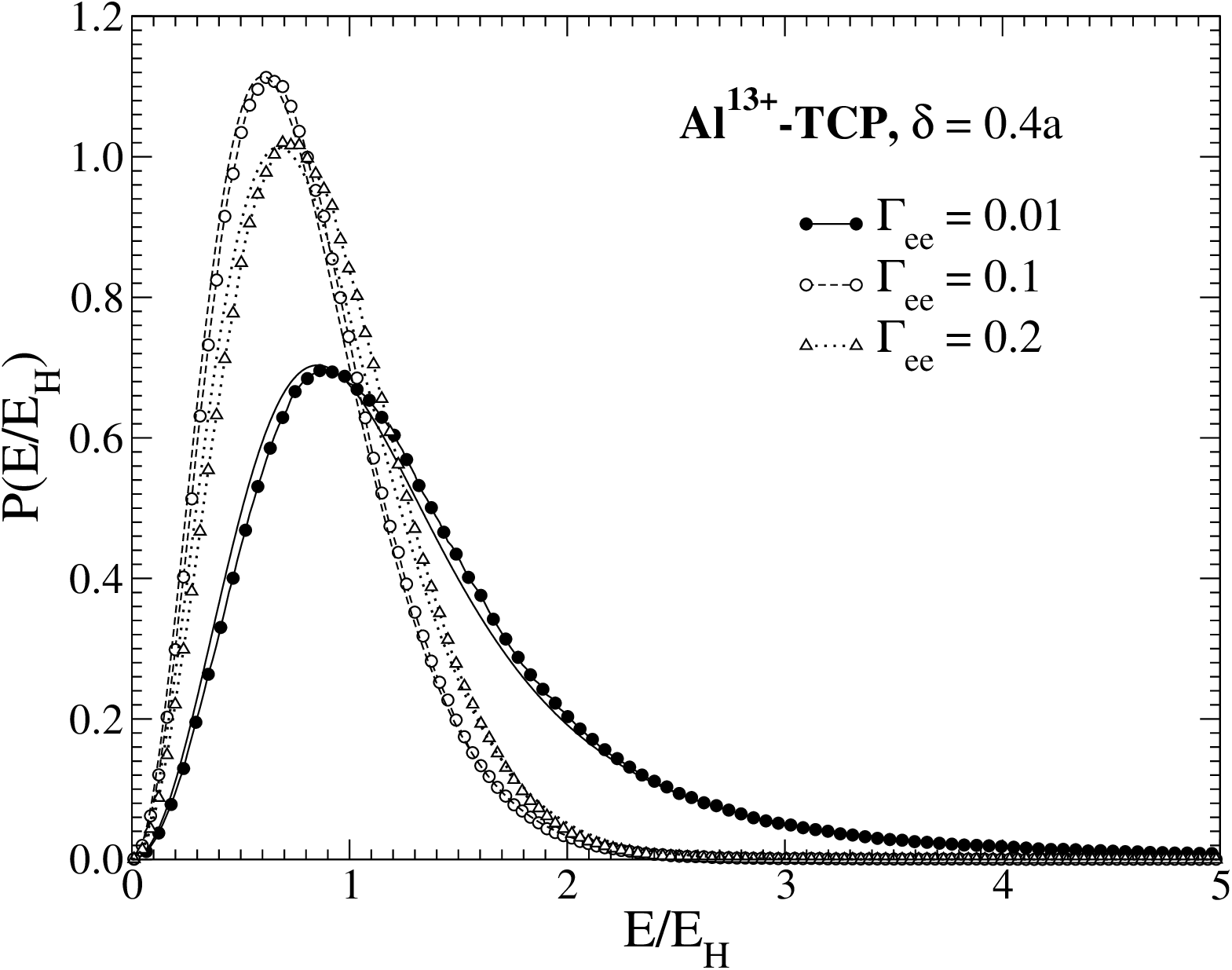}
\caption{Same as Fig.~12 with ${\bar\delta}=0.4$ and $\Gamma_{ee}=0.01$ (solid
lines),~$0.1$ (dashed lines), and~$0.2$ (dotted lines).}
\end{figure}

\begin{figure}
\includegraphics[width=8cm]{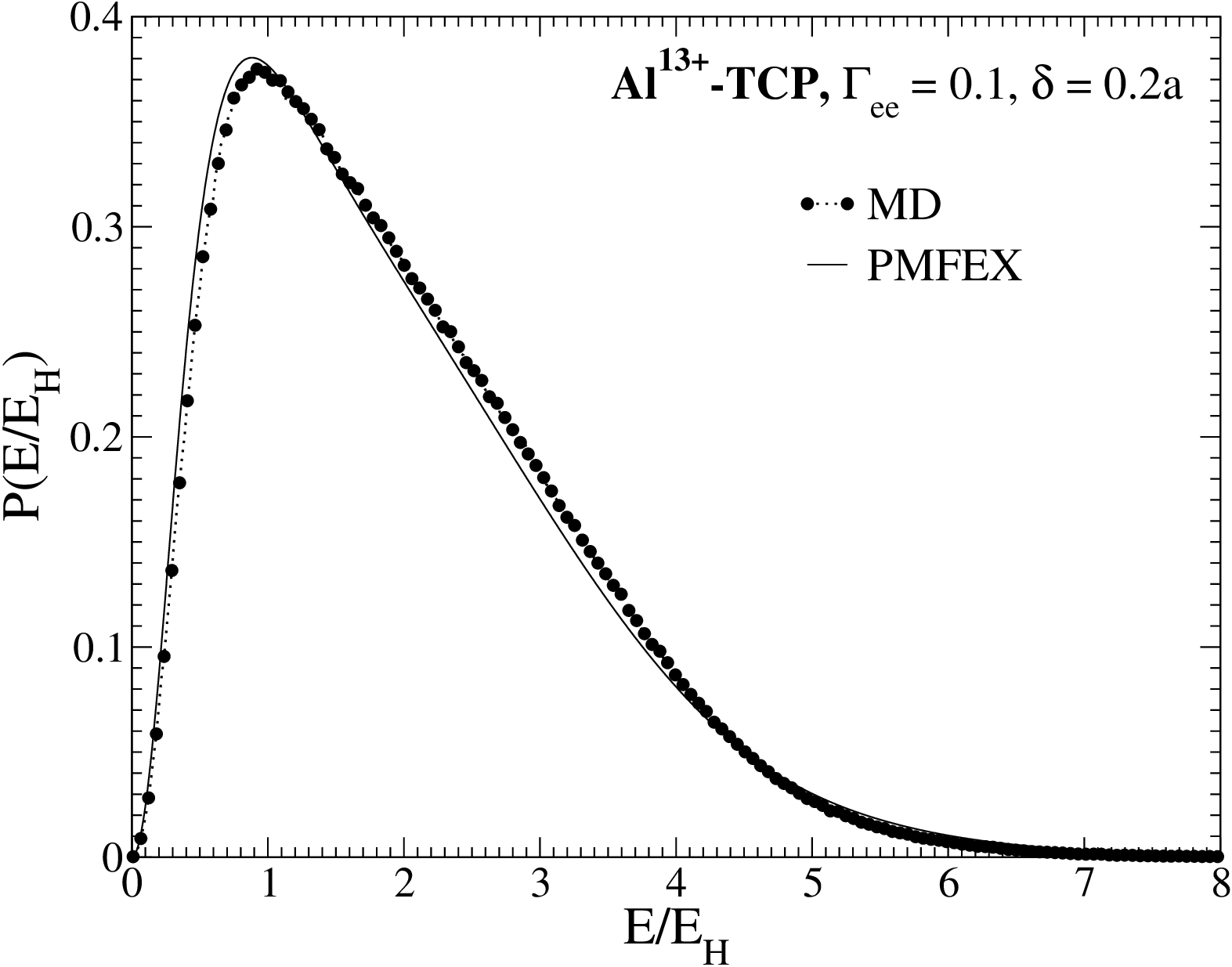}
\caption{Same as Fig.~12 for ${\bar\delta}=0.2$ and $\Gamma_{ee}=0.1$.}
\end{figure}

\begin{figure}
\includegraphics[width=8cm]{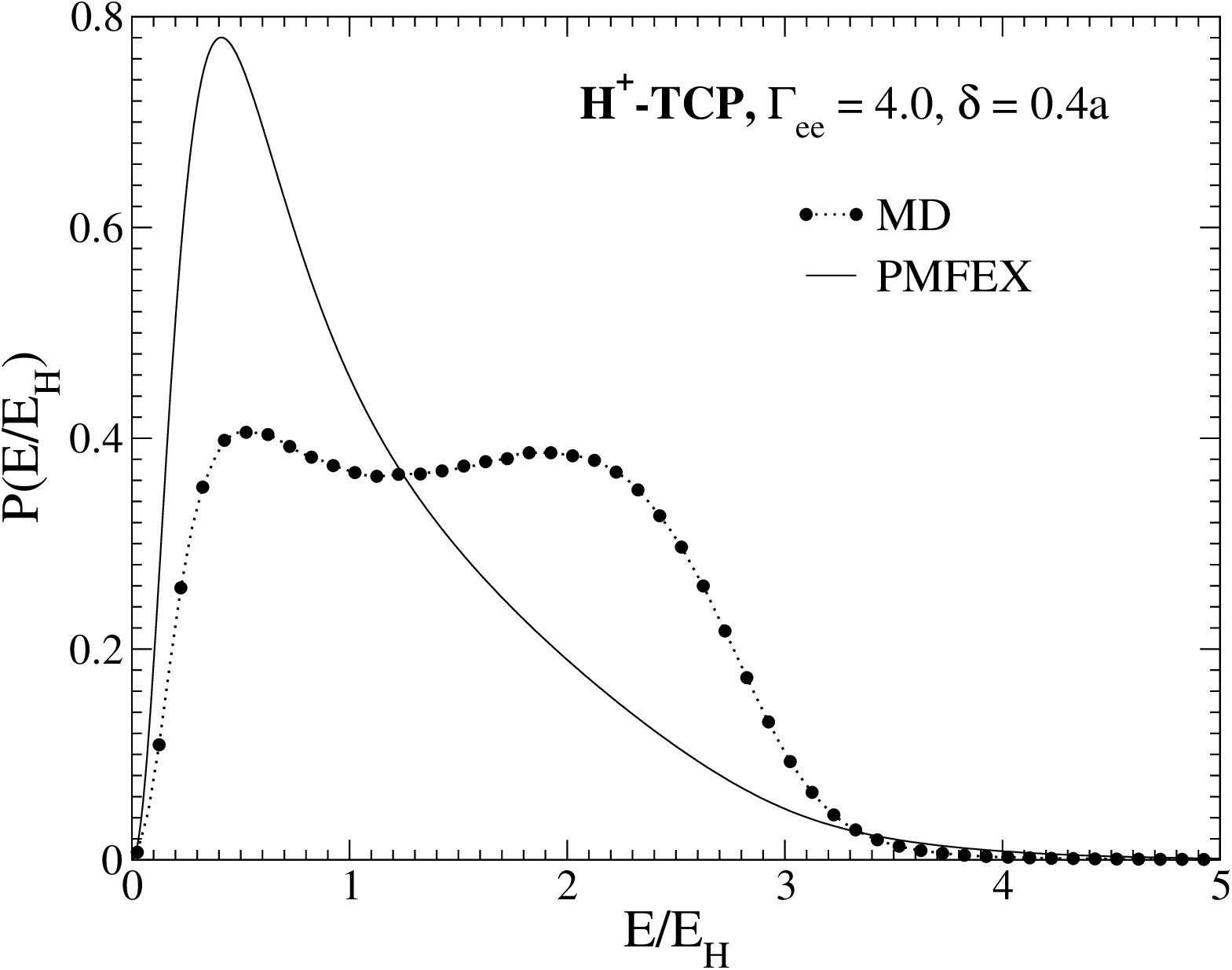}
\caption{Same as Fig.~11 for $\Gamma_{ee}=4.0$ and ${\bar\delta}=0.4$.}
\end{figure}

\end{document}